\documentclass{aastex631}

\def\kms{\hbox{km$\;$s$^{-1}$}}

\newcommand{\blos}{$\mathrm{B}_{\mathrm{LOS}}$}

\usepackage{graphicx,animate}
\usepackage{float}
\usepackage{booktabs}
\usepackage{soul}
\graphicspath{{images/}}
\usepackage{amsmath}
\usepackage{hyperref}
\begin{document}
\title{Thermodynamic Evolution of Plumes}
\author[0009-0002-8422-4310]{Biswanath Malaker}
\affiliation{Inter University Centre for Astronomy and Astrophysics, Pune, India - 411007}
\author[0000-0002-9253-6093]{Vishal Upendran}
\affiliation{Bay Area Environmental Research Institute, NASA Research Park, Moffett Field, CA, USA - 94035}
\affiliation{Lockheed Martin Solar and Astrophysics Laboratory, Palo Alto, CA, USA - 94304}
\author[0000-0003-1689-6254]{Durgesh Tripathi}
\affiliation{Inter University Centre for Astronomy and Astrophysics, Pune, India - 411007}

\begin{abstract}
Plumes are considered to play an important role in the origin of solar wind. However, an understanding of their thermodynamic evolution is not complete. Here, we perform a detailed study of a plume inside a coronal hole throughout its lifetime, using the observations from the Atmospheric Imaging Assembly (AIA) and the Helioseismic and Magnetic Imager (HMI). We find that the plume's formation is preceded by frequent occurrences of small-scale jets and jet-lets at its base, leading to the gradual development of plume haze. The plume rapidly developed within the first six hours into its well-known morphology. Light curves from all EUV channels exhibit a similar profile, suggesting its multi-thermal nature and intensity modulation over its lifespan. Moreover, the photospheric magnetic field dynamics at the plume's base are highly correlated with its light curve in 171~{\AA}. We calculate outflow velocities, observed prominently in the 171~{\AA} passband and mildly in the 193~{\AA} and 211~{\AA} passbands, with median speeds lower in higher temperature bands but occasionally comparable to the respective sound speeds. When data is averaged over larger spatial scales, the plume appears iso-thermal along its length, with constant temperature throughout its lifetime. However, an analysis of the differential emission measure at full resolution reveals the presence of higher-temperature plasma, indicating internal temperature structures within the plume. These results provide new insights into the formation, dynamics, and thermal properties of coronal plumes, placing tighter constraints on models to understand their thermodynamic evolution and potential role in the solar wind.

\end{abstract}

\keywords{Solar Coronal Plume --- Coronal Hole --- Solar Coronal Jets --- Plumelets -- Jetlets }

\section{Introduction} \label{sec:intro}
Solar coronal plumes are elongated, long-lasting plasma structures within Coronal Holes (CHs). Plumes are particularly conspicuous within the polar coronal holes due to their enhanced density with respect to the background. They are distributed inside the CHs at all latitudes \citep{wang-muglach-2008}. Plumes have been observed in observations taken across the electromagnetic spectrum, i.e., in white light, Extreme Ultraviolet (EUV), and X-rays \citep{SDO1,boe_2020_wlplume,cho_2024_wlplume} and are suggested to play an important role in the supply of mass and energy to the solar wind. Plumes are known to be slightly cooler than the inter-plume regions~\citep{wilhelm}, typically measuring around 1 MK.
Plumes are proposed to form due to reconnection between newly emerged magnetic bipoles with pre-existing and effectively open magnetic field lines \citep{wang-1998}. In this model, plumes form due to the evaporation of chromospheric plasma that is heated by the impulsive deposition of heat from magnetic reconnection, similar to the formation of jets \citep{Shibata, Yokoyama, Yokoyama-1}. Recent observations indicate a clear association between plumes and coronal jets \citep[see, e.g.][]{Raouafi-2009, Pucci-2014, raouafi-stenborg-2014, Kumar-2022}. However, plumes exhibit a long lifetime, averaging $\approx20$ hours to sometimes persisting for up to a week \citep{poletto-2015}, whereas jets in the CHs last only for about 1 to 30 minutes \citep{Raouafi-2016}. Observations suggest the existence of tiny jetlets in plumes, which are seen to occur very frequently, possibly resulting from magnetic flux cancellation between small concentrations of minority polarity with the dominant polarity~\citep[see e.g.][]{raouafi-stenborg-2014}. 

\cite{uritsky-2021} showed that plumes contain several filamentary structures  (called plume-lets) and established a positive correlation between the number of plume-lets and the brightness of the plume at a given moment. This correlation was observed to saturate at the peak of the plume's brightness. Such internal structure raises questions on whether plumes consist of small-scale threads, curtain-like shapes, or beam-like \citep{gabriel-2009}. 

Plumes exhibit specific brightness modulation throughout their lifetime, with a tendency to reappear in the exact location where they had previously faded out \citep{de-forest-2001a, Pucci-2014}. This phenomenon poses a challenge in defining the exact duration of a plume's lifespan. \cite{raouafi-stenborg-2014} observed that their plume of study faded away when the dominant magnetic field concentration at the plume's base dispersed. \cite{Pucci-2014}, on the other hand, observed that their plume of study started disappearing when the bright point at the base of the plume started to get dispersed and faint. Therefore, it is important to understand the thermal nature of plumes. Specifically, we ask: Do the plumes fade away from one imaging channel and reappear in another? Do they have internal multi-thermal structures? Do they have any dependence on the underlying magnetic field?

In this paper, we perform an in-depth study of the thermodynamic evolution of one plume from its birth to its disappearance, with the explicit aim of addressing the questions as mentioned above relating to the formation and evolution of a plume, including its internal structure. The rest of the paper is structured as follows. In \S.~\ref{Data Processing}, we describe the data used and the reduction process. We then investigate the plume through its three main phases: appearance, peak, and disappearance in \S.~\ref{Intensity and magnetic field evolution}~\&~\S.~\ref{sec:light_curve_and_flux_density}. In \S.~\ref{Outflows}, we analyze the plume through time-distance analysis, and in \S.~\ref{DEM Structure}, through Differential Emission Measure (DEM). We summarize and discuss our results in \S.~\ref{summary} and finally conclude in \S.~\ref{conclusions}.
\section{Observations and Data} \label{Data Processing} 

\begin{figure}[!]
    \centering
    \includegraphics[width=\linewidth]{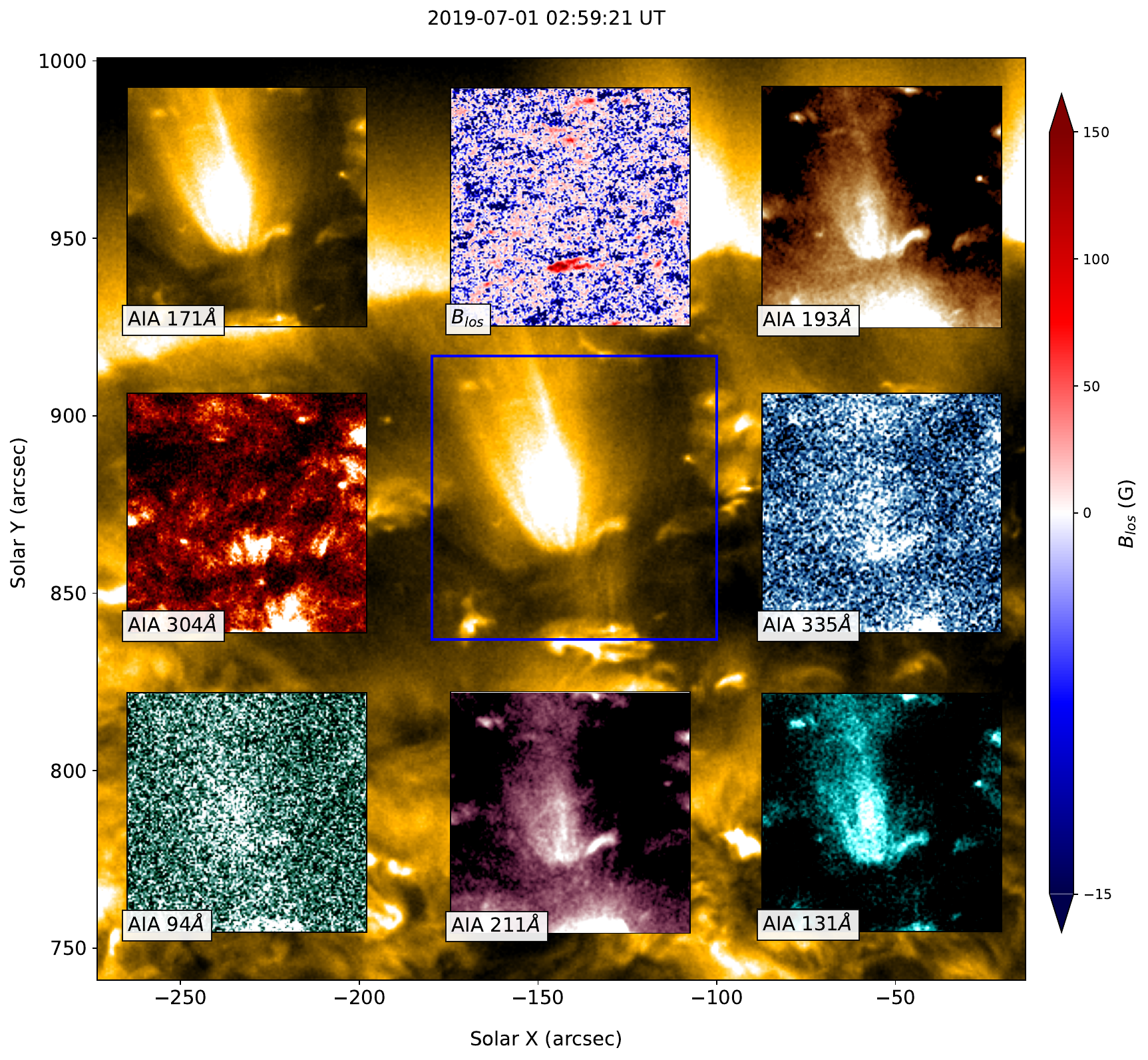}
    \caption{Background: The plume at its peak intensity as observed in 171~{\AA} band of AIA on July 1, 2019, at 02:59:21~UT. Inset: Plume in all AIA passbands observed when it peaked in 171~{\AA}, along with the {\blos} produced by HMI in the top middle inset. The FOV of the insets corresponds to the background blue box. The color bar is provided for the {\blos} in the inset.} 
    \label{fig:broad_fig}
\end{figure}
\begin{figure}[htpb!]
    \centering
        \includegraphics[width=\textwidth]{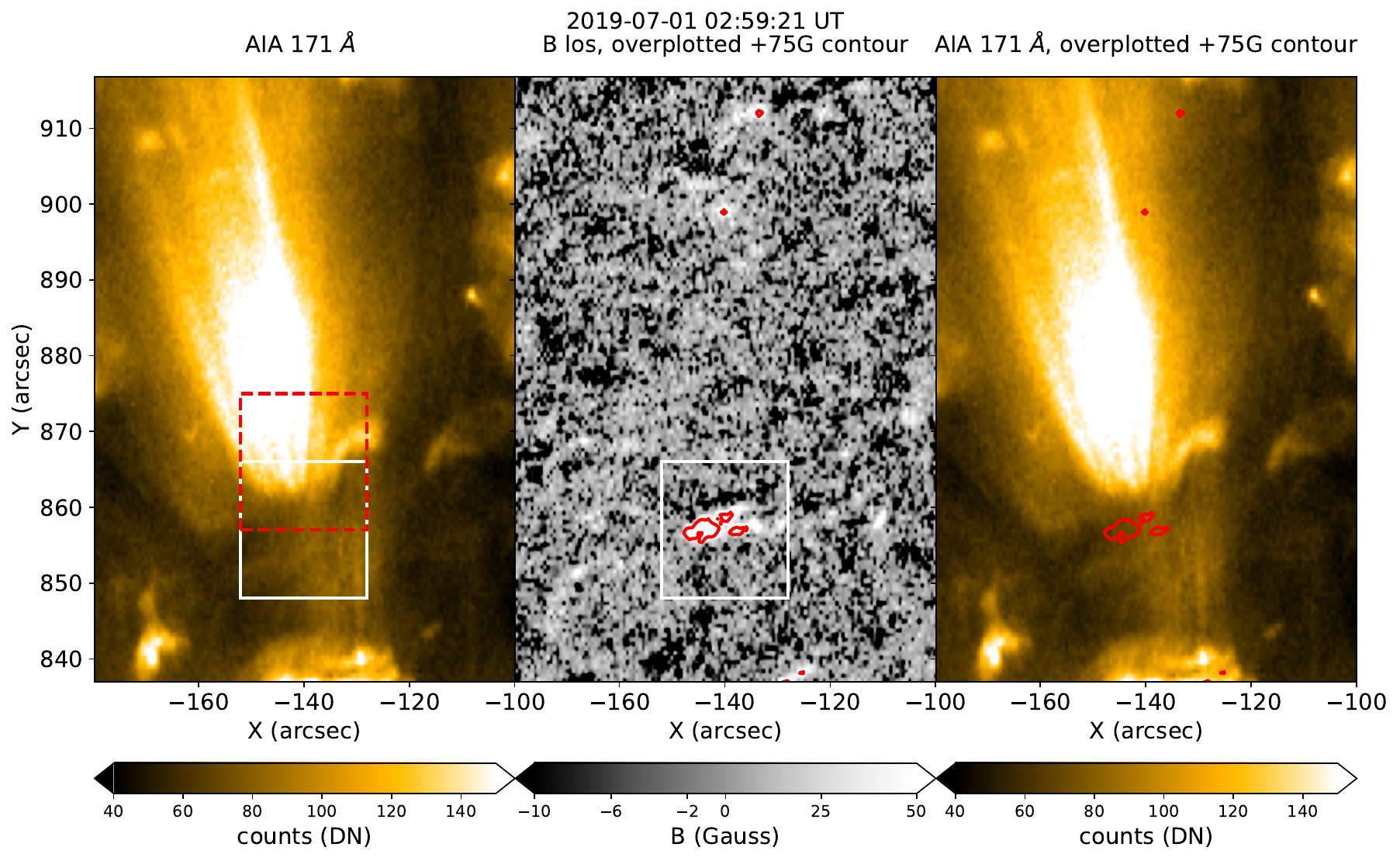}
    \caption{Left: The plume at the peak intensity observed by AIA 171~{\AA} on July 01, 2019. The overplotted red and white boxes are to obtain the intensity light curve and to study the evolution of magnetic flux density. Middle: The corresponding line of sight magnetogram, over-plotted with +75~G red contours. The white solid box (identical to what is shown in the left panel) is used to obtain magnetic field evolution. Right: The plume seen in AIA 171~{\AA} with over-plotted +75G magnetic field contour in red. } 
    \label{fig:light_box_positions}
\end{figure}
In this work, we use data from Atmospheric Imaging Assembly~\citep[AIA;][]{AIA1} and Helioseismic and Magnetic Imager~\citep[HMI;][]{HMI1} onboard the Solar Dynamics Observatory~\citep[SDO;][]{SDO1}. AIA records full disk images of the Sun in 7 EUV passbands centered at 94~{\AA}, 131~{\AA}, 171~{\AA}, 193~{\AA}, 211~{\AA}, 304~{\AA}, and 335~{\AA}, with a time cadence of $\approx$12~s and pixel size of $\approx$0.6{\arcsec}. Using these passbands, AIA covers a broad range of temperatures from $6\times10^{4}$K to $2\times10^{7}$K, with different passbands being sensitive to different temperatures under varying activities \citep{O'Dwyer-2010}. To study the structure of photospheric magnetic flux at the base of plumes, we have used the line-of-sight (LOS) magnetograms from HMI that provide full disk LOS magnetograms at the temporal resolution $\approx$45~s with a pixel size of $\approx$0.5{\arcsec}. To obtain the level 1 data from AIA and HMI, we have used the JSOC cut-out service\footnote{\href{http://jsoc.stanford.edu/}{http://jsoc.stanford.edu/}}. 

For this study, we selected a plume positioned around (-147\arcsec, 857\arcsec) on 2019-07-01 03:00 UT. In Fig.~\ref{fig:broad_fig}, we display the plume at its peak brightness in 171~{\AA}, on 2019-07-01 at 02:59:21~UT. The insets display the plumes observed in other passbands, along with {\blos} measurements from HMI. The field of view (FOV) covered by the insets corresponds to the region bounded by the blue box in the main frame. Note that the plume is visible in all the channels except 94~{\AA}, 304~{\AA}, and 335~{\AA}. While in 304~{\AA}, there is strong emission at the base of the plume, in 335 and 94, the structure appears hazy and noisy. The LOS magnetogram clearly shows the presence of a dominant clump of positive polarity magnetic field at the base of the plume.

\begin{figure}[ht!]
    \centering
    \includegraphics[width=\linewidth]{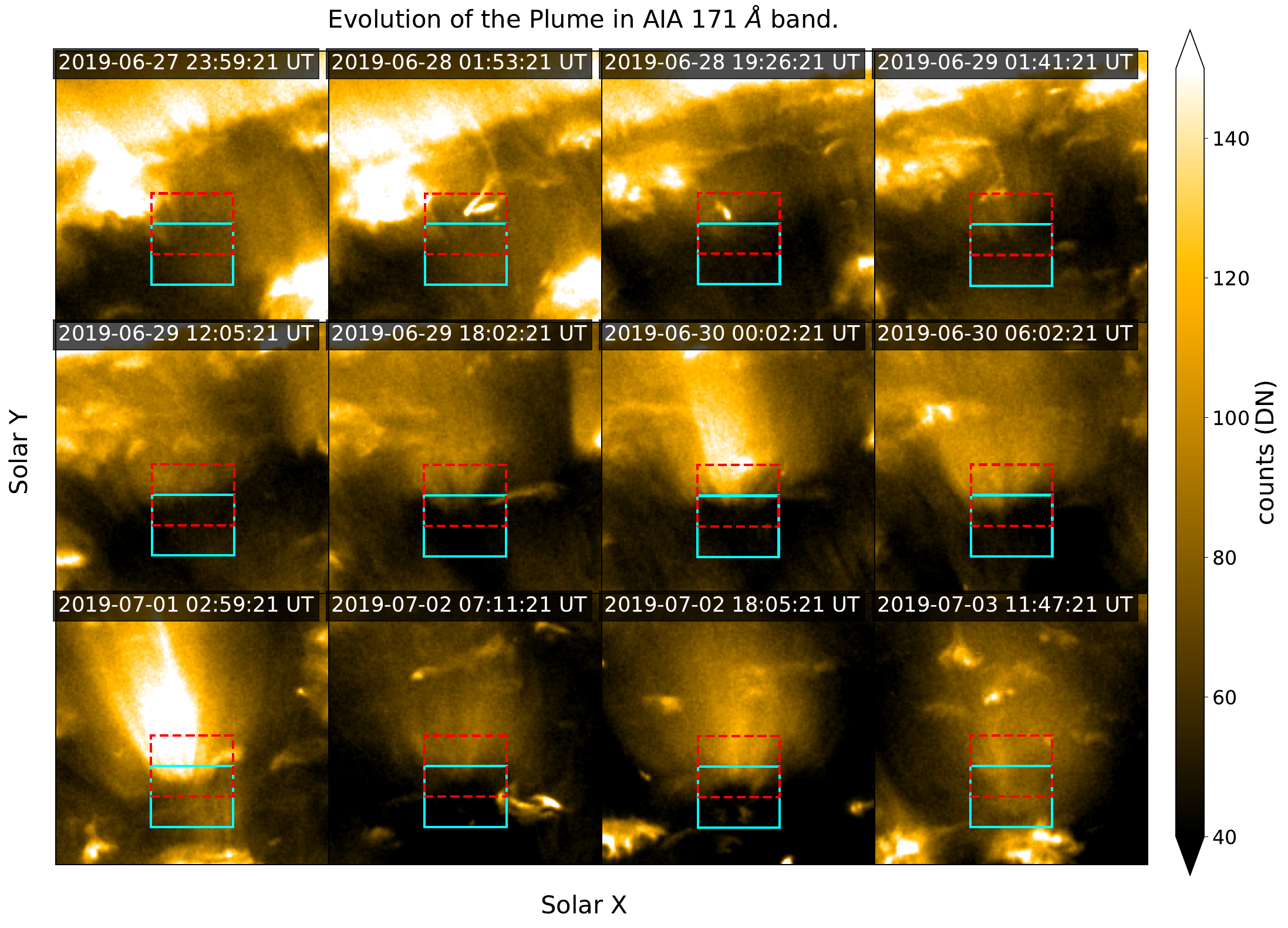}
    \caption{Evolution of the plume from its first appearance to final disappearance as observed by AIA 171~{\AA}. The lower cyan box (solid) estimates magnetic flux from the HMI line-of-sight magnetogram, and the upper red box (dashed) estimates the brightness of the plume in different AIA/EUV passband. Time has been provided with corresponding plots.}
    \label{fig:plume_evolution_171}
\end{figure}
\section{Data analysis and results} \label{Methodology and Results and Discussions}
\subsection{General intensity and magnetic field evolution} \label{Intensity and magnetic field evolution}
We first study the long-term evolution of the plume from its birth (2019-06-28 00:00 UT) to its disappearance (2019-07-04 00:00 UT). We perform a running average of 5 AIA images taken approximately every 12 sec (effective cadence $\approx1$ minute) to improve the signal-to-noise ratio. However, we did not perform any averaging process with the HMI data.

In the left panel of Fig.~\ref{fig:light_box_positions}, we display a portion of the disk showing the plume as recorded by AIA 171~{\AA} on 2019-07-01 at 02:59:21 UT. The middle panel shows {\blos} corresponding to the left panel. The over-plotted red contours correspond to the magnetic flux density of +75~G. The right panel is the same as the left panel but over-plotted with the same contours as shown in the middle panel. We note that the cluster of +75~G flux density is located at the base of the plumes. In the left panel, we select two boxes of width 24{\arcsec} and height 18{\arcsec}. The lower box (solid white boundary) is used to study the evolution of the magnetic field, while the upper box (red dashed boundary) is used to study the evolution of the plume brightness in different AIA EUV bands.

In Fig.~\ref{fig:plume_evolution_171}, we show the evolution of the plume in the 171~{\AA} band from its first appearance to disappearance. Time increases from left to right and from top to bottom. The red dashed and cyan solid boxes are the same as the red dashed and white solid boxes in Fig.~\ref{fig:light_box_positions}. In the top left image of Fig.~\ref{fig:plume_evolution_171}, the red box encloses a region inside the coronal hole devoid of any plume. In the following two images, we observe the launch of two jets, leading to the appearance of a plume haze in the fourth image. This plume haze stays in the first image of the middle row, transforming into the well-known morphology of the plume in the subsequent two images. We then observe a rapid growth in the plume intensity from $\approx$18:00 UT on 2019-06-29, which starts to fade from $\approx$00:00 UT to $\approx$06:00 UT on 2019-06-30. We observe several such modulations in the plume intensity throughout its lifetime. On 2019-07-01 at $\approx$03:00 UT, the plume is at its peak brightness, as seen in the left image of the last row. It fades away around 2019-07-02, $\approx$07:00 UT, and reappears, as seen in the following two images. Finally, the bottom-right image shows the plume haze just before its final disappearance.

\begin{figure}[hb!]
    \centering
    \includegraphics[trim=0 2cm 0 0, width=\linewidth]{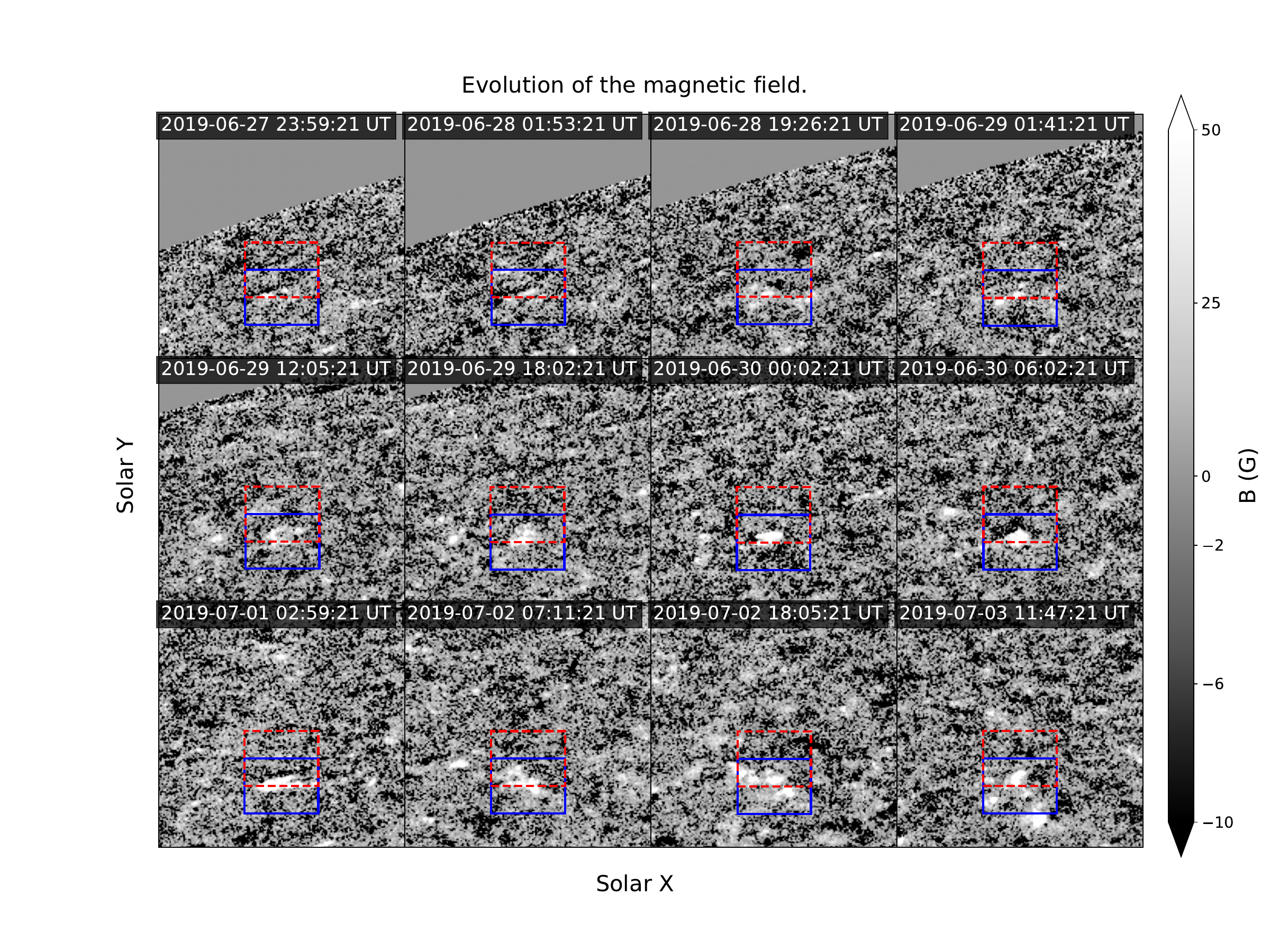}
    \caption{Evolution of line of sight magnetic flux density at the base of the plume as observed in HMI line of sight magnetograms. The lower blue box estimates magnetic flux from the HMI line-of-sight magnetogram, and the upper red box estimates the brightness of the plume in different AIA/EUV passband. Time has been provided with corresponding plots.}
    \label{fig:plume_evolution_magnetic}
\end{figure}
\begin{figure}[htpb!]
    \centering
        \includegraphics[width=\linewidth]{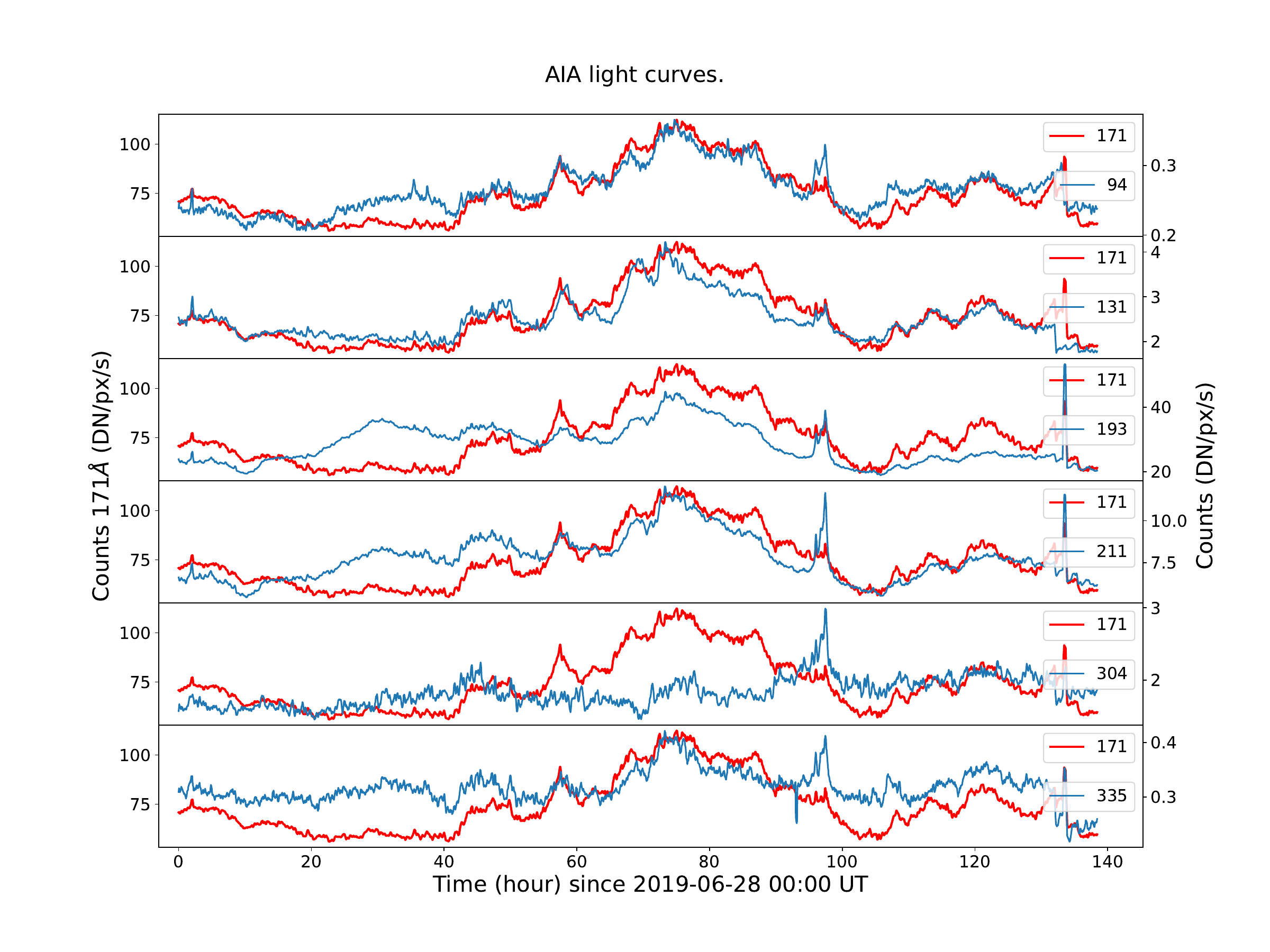}
    \caption{Intensity light curves in all AIA passbands (in blue) were obtained by averaging over the red boxed region shown in Fig.~\ref{fig:plume_evolution_171}. The reference light curve from the 171~{\AA} passband is shown in red. All the curves are smoothed using a running boxcar of 15 min.}\label{fig:light_curves}
\end{figure}

In Fig.~\ref{fig:plume_evolution_magnetic}, we display the {\blos} magnetograms corresponding to the AIA 171~{\AA} maps shown in Fig.~\ref{fig:plume_evolution_171}. The magnetic flux maps show the presence of a weak positive dominant magnetic field at the base of the plume that gradually increases with time. At the peak of the plume's brightness, the dominant positive magnetic field strength also reaches its peak, reaching a value of around +200~G. Finally, during the disappearance of the plume, the emerging dominant magnetic fields disintegrate and scatter away.  

\subsection{Light curves of intensity and magnetic flux density analysis}
\label{sec:light_curve_and_flux_density}

We now study the AIA light curves averaged over the red dashed box in Fig.~\ref{fig:plume_evolution_171}. For this purpose, we consider the 1-minute averaged AIA data at a cadence of $\approx$3 minutes and average the intensities inside the red box. We present the light curves for each AIA passband in Fig.~\ref{fig:light_curves}. We also over-plot the 171~{\AA} light curve in each panel for reference. In each plot, the left y-axis corresponds to the 171~{\AA}, while the right y-axis corresponds to the other passband. The long-term modulation in plume brightness we noted in Fig.~\ref{fig:plume_evolution_171} may be seen in the light curve in Fig.~\ref {fig:light_curves}. The light curves show that all the passbands follow similar trends except 304~{\AA}. Although the plume is prominent in the AIA 171{~\AA} channel, it shows signatures in AIA 131{~\AA}, AIA 193{~\AA}, and AIA 211~{\AA}. We note a sharp enhancement in the intensity just before the 100th hour. This enhancement occurs due to the appearance of a newly formed loop within the FOV that is included in the boxed region, and it is not related to plume evolution. 

\begin{figure}[ht!]
    \centering
    \includegraphics[width=\linewidth]{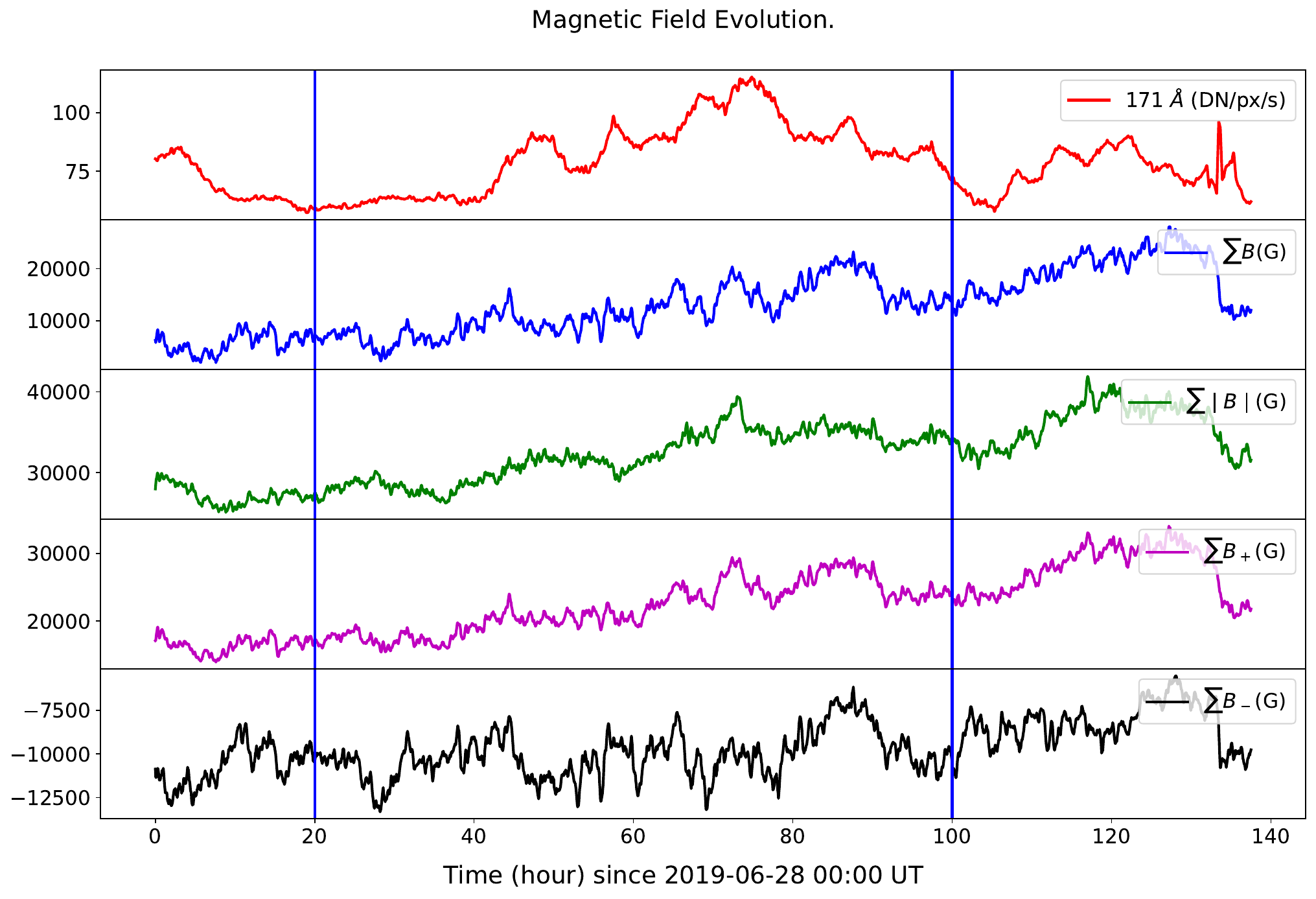}
    \caption{The evolution of magnetic flux density in the blue-boxed regions is shown in Fig.~\ref{fig:plume_evolution_171}. From top to bottom, the light curve obtained for AIA~171{\AA} in the red-boxed region (fig.~\ref{fig:plume_evolution_171}), the evolution of total unsigned flux density ($\Sigma B$), absolute magnetic flux density ($\Sigma \vert B\vert $), total positive ($\Sigma B_+$) and negative flux density ($\Sigma B_-$). All the curves are smoothed using a running boxcar of 15 min. The region between the two vertical lines has been selected to obtain the correlation between the positive magnetic field and the AIA 171~{\AA} light curve in Fig.~\ref{fig:171A_B} \& \ref{fig:correlation}.} \label{fig:magnetic_evolution}
\end{figure}

To obtain the evolution of the magnetic flux density, we consider the HMI {\blos} magnetogram with a cadence of 3 minutes, corresponding to the AIA observations. We construct the light curve by summing the flux in the cyan-boxed region shown in Fig.~\ref{fig:plume_evolution_magnetic}. In Fig.~\ref{fig:magnetic_evolution}, we show the 171~{\AA} light curve (top panel), total signed magnetic flux density ($\Sigma B$), total unsigned magnetic flux density ($\Sigma \vert B\vert $), total positive flux ($\Sigma B_+$), and total negative flux ($\Sigma B_-$) in different panels. The curves for $\Sigma B_+$ and $\Sigma B_-$ show an overall increasing trend that is also reflected in the $\Sigma B$ and $\Sigma \vert B\vert $. We note a substantial dip in $\Sigma B_-$ (bottom panel) just after 40 hrs and an increase again, albeit with many ups and downs till about 60 hrs. This dip is also reflected in the $\Sigma B$ (see second panel from top).

\begin{figure}[hb!]
    \centering
    \includegraphics[width=\linewidth]{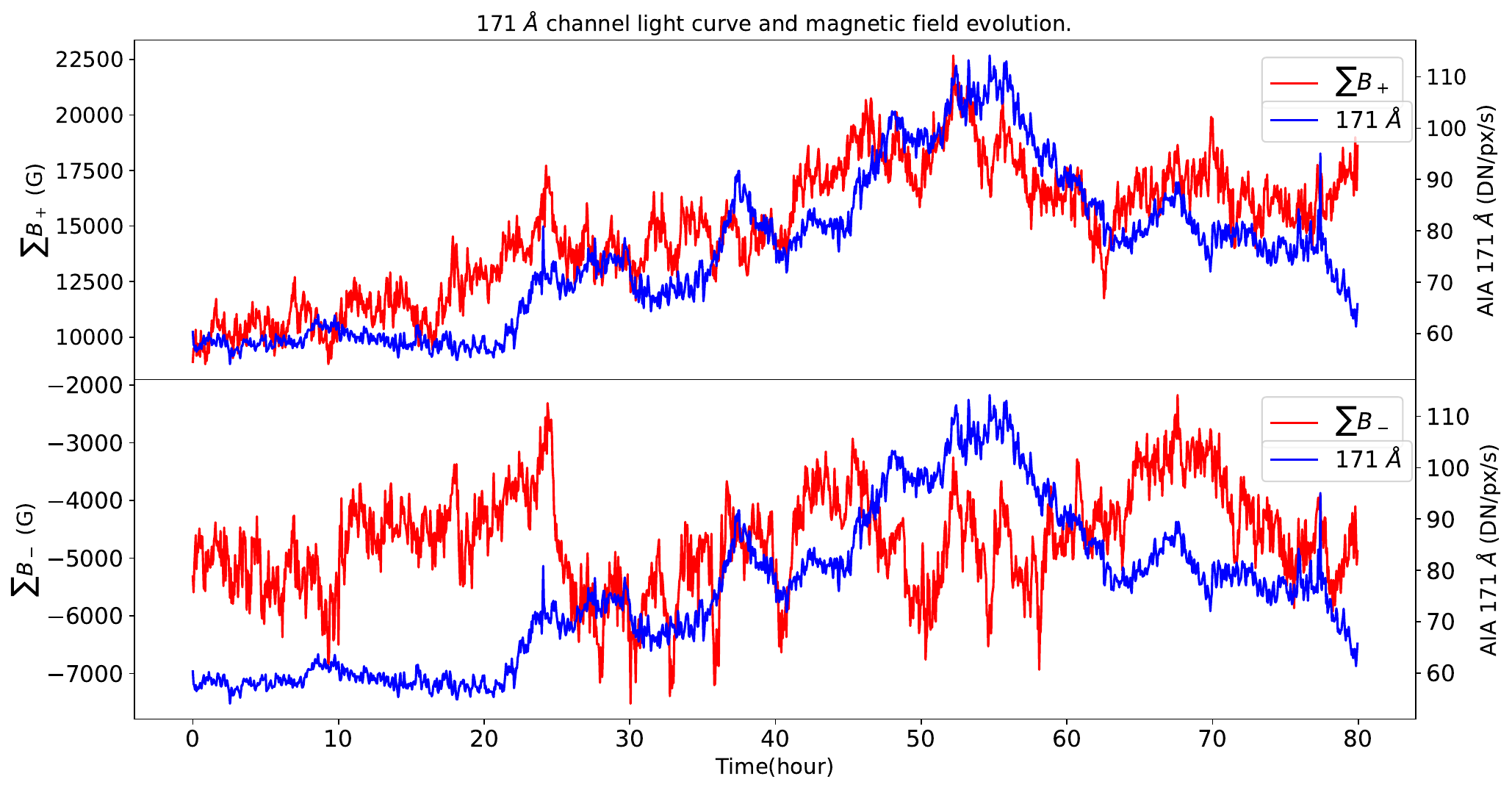}
    \caption{The red curve is the change in emerging positive magnetic flux ($\sum{B_+}$ at the top) and negative magnetic flux ($\sum{B_-}$ at the bottom) with time, and the blue curve is the AIA 171 {\AA} light curve. These quantities are subsets of the Figs.~\ref{fig:light_curves},~\ref{fig:magnetic_evolution} without smoothing. The plot is obtained for data between 2019-06-28 20:00:00 UT to 2019-07-02 04:00:00 UT (enclosed between the two vertical lines), and no running box car average has been performed, Unlike figs~\ref{fig:light_curves} and ~\ref{fig:magnetic_evolution}.}
    \label{fig:171A_B}
\end{figure}

\begin{figure}[htpb!]
    \centering
    \includegraphics[width=\linewidth]{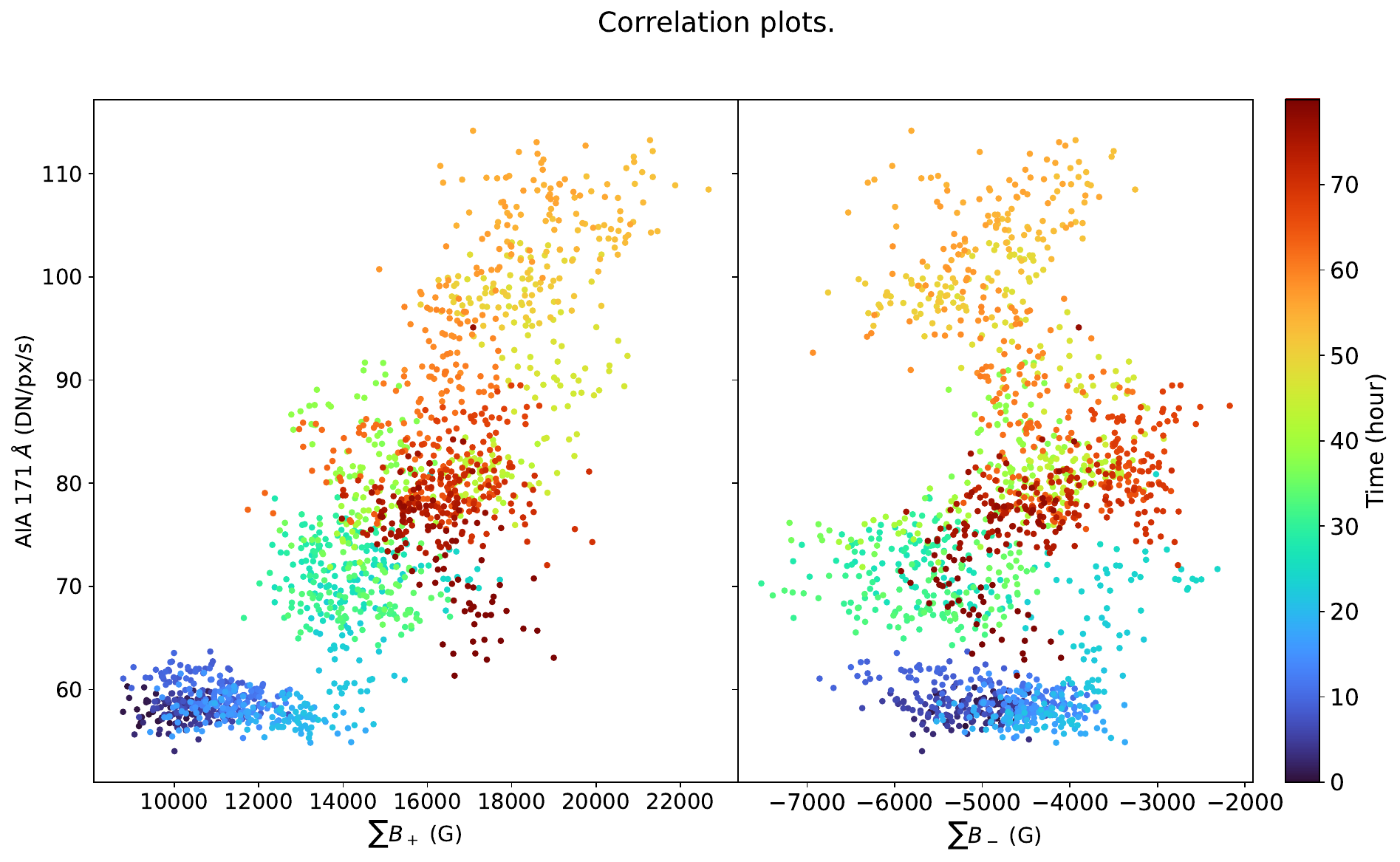}
    \caption{Correlation between the 171~{\AA} channel light curve with $\sum{B_+}$ (left) and $\sum{B_-}$ (right) at the plume's base.  Colors indicate time in hours, with blue at the start of the activity, green/yellow towards the peak of the activity, and red at the end of the activity.}
    \label{fig:correlation}
\end{figure}

To study the evolution of plume at the highest cadence, we study the variation of AIA 171 light curve and photospheric magnetic flux for the duration enclosed by the two vertical lines in Fig.~\ref{fig:magnetic_evolution}, in Fig.~\ref{fig:171A_B}. The AIA light curve is shown in a blue curve, while the positive and negative magnetic flux densities are over-plotted in red. Clearly, the two curves show similar temporal behavior by eye. To quantify the causal relation between $\Sigma B_+$, $\Sigma B_-$, and AIA 171~{\AA} intensity, we construct a scatter plot of the magnetic flux with 171~{\AA} intensity in Fig.~\ref{fig:correlation}. The scatter plot is color-coded with the time elapsed since 2019-06-28, 20:00:00 UT. Blue corresponds to the start of the jetting activity, green/yellow/orange to the peak, and red to the end of the jetting activity. This representation may be considered as a state-space diagram with the system moving about the AIA 171~{\AA} and magnetic flux space in time. We display the diagram with $\Sigma B_+$ on the left and $\Sigma B_-$ on the right panels of Fig.~\ref{fig:correlation} respectively.

From the left panel of Fig.~\ref{fig:correlation}, we see that the changes in the $\Sigma B_+$ are not correlated with plume intensity at initial times. However, with time, a linear association builds up between $\Sigma B_+$ and plume intensity, with the intensity increasing with $\Sigma B_+$. Towards the end of the plume, the intensity drops and stays nearly constant, as does $\Sigma B_+$. The Pearson correlation coefficient computed for all the points is $\approx$0.83. This high correlation is attributed to the long-term variation of the plume brightness rather than short-term brightness modulation. Hence, we find the plume exists as long as the dominant magnetic field is present at the plume's base and its intensity is proportional to the dominant magnetic field.

In the right panel of Fig.~\ref{fig:correlation}, we see that the changes in the $\Sigma B_-$ are once again not correlated with plume intensity initially. However, with passing time, we see first a rise in the negative flux, with almost no changes in the plume intensity (cluster of green points). Interestingly, the negative flux starts reducing after that, suggesting magnetic flux cancellation, with a corresponding rise in the plume intensity (green to yellow points). After this rise, we find multiple episodes of increase and reduction in $\Sigma B_-$, while the plume intensity rises to its peak and remains constant. Towards the end of the jetting period, the intensity drops while $\Sigma B_-$ goes approximately to pre-jet values. Thus, we find correspondence between an increase in jetting activity and the cancellation of minority polarity and possible sustenance of the plume through cancellation mechanisms.


\subsection{Outflows}\label{Outflows}

\begin{figure}[htpb!]
    \centering
        \includegraphics[scale=0.5]{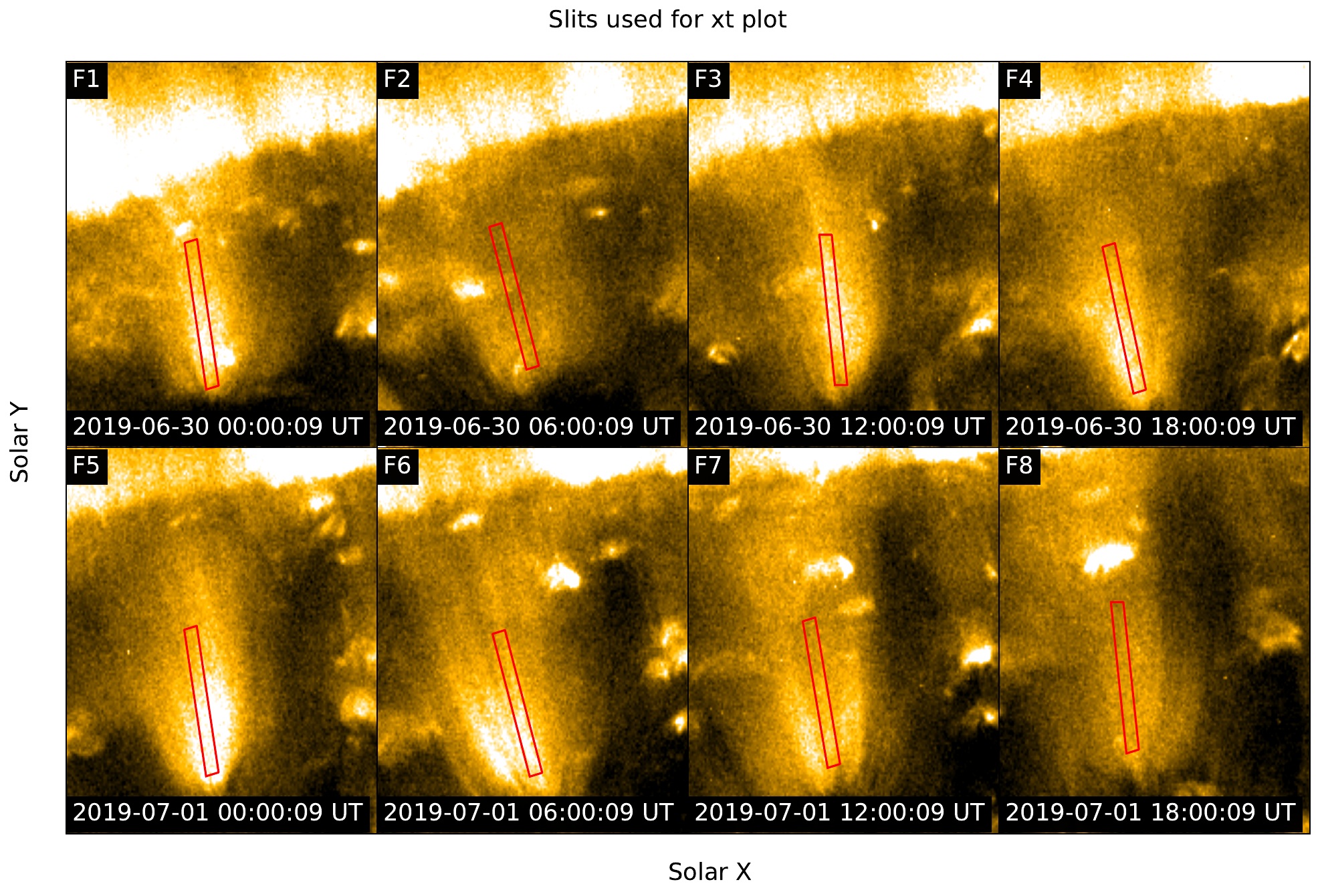}
    \caption{AIA~171~{\AA} images overplotted with slits considered for deriving the space-time plots. The corresponding start times of the xt-plots are shown in the plots.}\label{fig:xt_slits}
\end{figure}

We now analyze plasma flows inside the plume using space-time plots. We consider eight different time windows during the peak of the plume brightness: \textit{viz.} F1 to F8 as marked in Fig.~\ref{fig:xt_slits}, each spanning two hours and separated by six hours, starting from 2019-06-30 00:00 UT and ending at 2019-07-01 18:00 UT. Representative timestamps from each of these windows are shown in  Fig.~\ref{fig:xt_slits}. To create the space-time plot, we chose a slit width of $5\arcsec$ and a height of $40\arcsec$ as shown in Fig.~\ref{fig:xt_slits}. We study the space-time plot using running difference on 12 s cadence data recorded using 171~{\AA}, 193~{\AA}, and 211~{\AA}. In addition, we have smoothed the running difference in time with a 1-dimensional Gaussian filter of width $\approx$24 sec.

We display four of the space-time plots obtained at different time instances in Figs.~\ref{fig:xt_plot12}~\&~\ref{fig:xt_plot34}, while others are put in the Appendix in Figs.~\ref{fig:xt_plot56}~\&~\ref{fig:xt_plot78}. The top (bottom) panel of Fig.~\ref{fig:xt_plot12} corresponds to the time when the plume undergoes its first brightening (dimming), on June 30th, starting at 00:00~UT (06:00~UT). Similarly, in Fig.~\ref{fig:xt_plot34}, we show the space-time plot when the plume was at its peak (end state) on July 01st, starting at 00:00~UT (18:00 UT). In both Figs.~\ref{fig:xt_plot12}~\&~\ref{fig:xt_plot34}, we show the space-time plots for 171~{\AA}, 193~{\AA}, 211~{\AA} in top, middle and lower rows, respectively.

From these space-time plots, the presence of bright ridges next to the black ridges is considered to be the signature of outflows. In all the space-time plots, the bright ridges are most prominent in 171~{\AA}, followed by 193~{\AA} and then 211~{\AA}. It is also apparent that the propagation is detected to the highest heights in 171 and lowest in 211~{\AA}. Comparison between Figs.~\ref{fig:xt_plot12}~\&~\ref{fig:xt_plot34} suggests that there are far more flows during the peak of the plume, while the number of peaks is the lowest during the first dimming (bottom panel of Fig.~\ref{fig:xt_plot12}) and the end phase (bottom panel of Fig.~\ref{fig:xt_plot34}). There are also no clear velocity signatures in 211~{\AA} during the early, first dimming, and end phases, while small signatures are seen during the peak of the plume. We also see the highest number of velocity signatures in 193~{\AA} during the peak of the plume. During the peak intensity of the plume, the presence of velocity signatures in all three channels suggests that the plasma is multi-thermal.

\begin{figure}[htpb!]
    \centering
    \includegraphics[width=\linewidth]{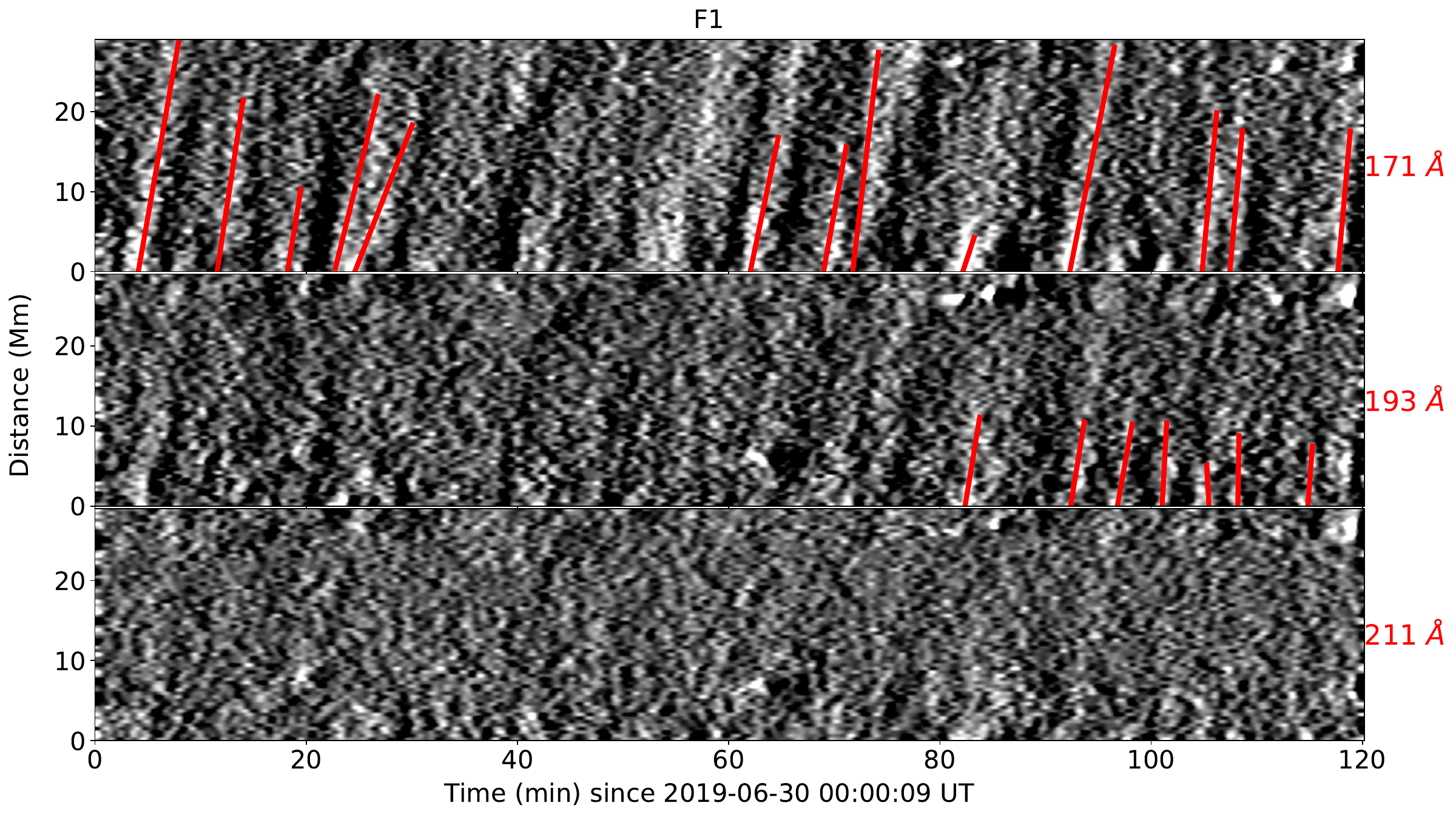}
    \rm{I}. First enhancement of the plume.
    \includegraphics[width=\linewidth]{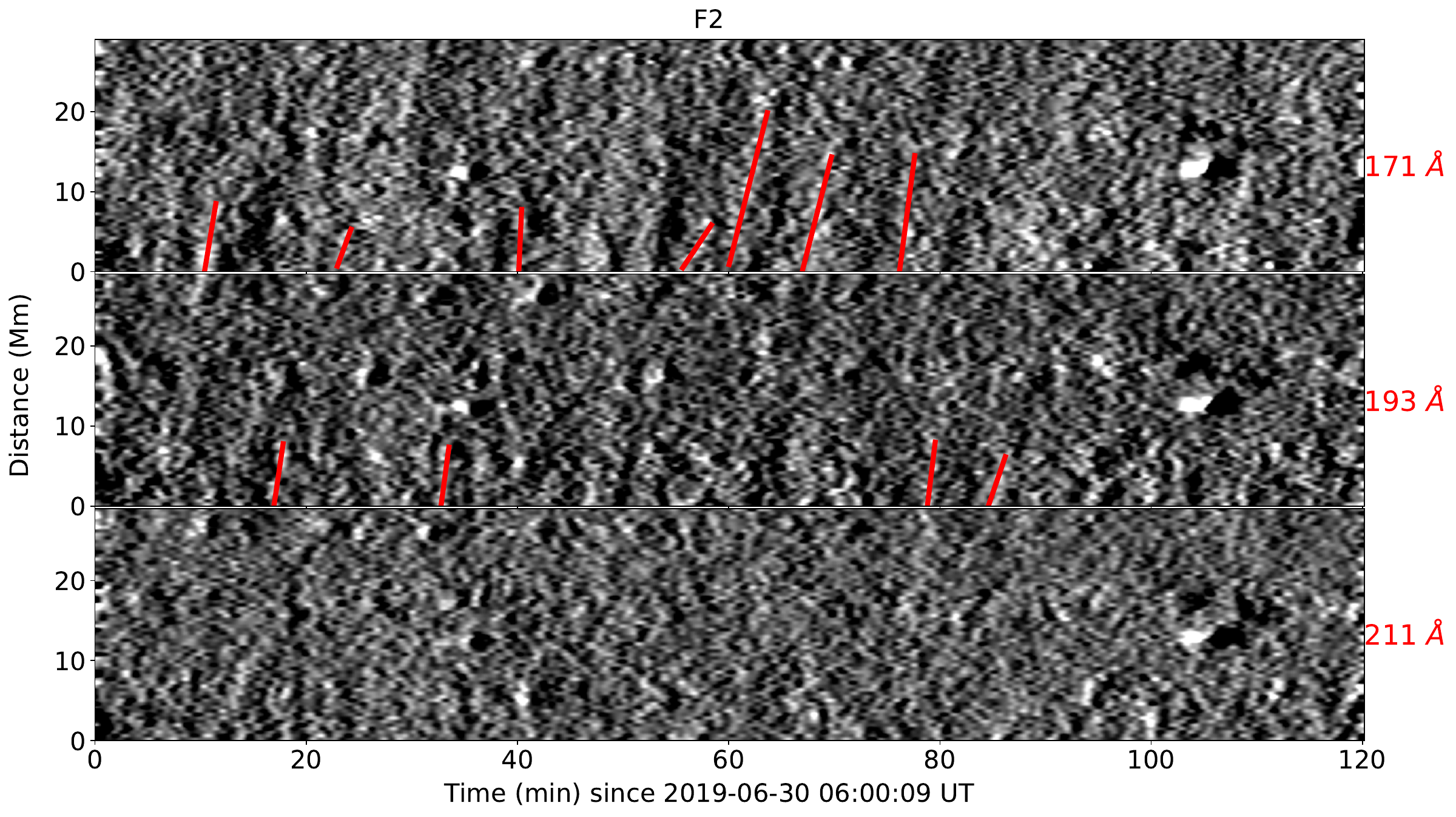}
    \rm{II}. First dimming of the plume.
    \caption{Panel \rm{I}: Space-time plot during the first brightening. Panel \rm{II}: Space-time plot at the first dimming of the plume. From top to bottom for each panel, we display the plot in 171~{\AA},193~{\AA}, and 211~{\AA} channels. Red lines are drawn to show outflow lines.}
    \label{fig:xt_plot12}
\end{figure}
\begin{figure}[htpb!]
    \centering
    \includegraphics[width=\linewidth]{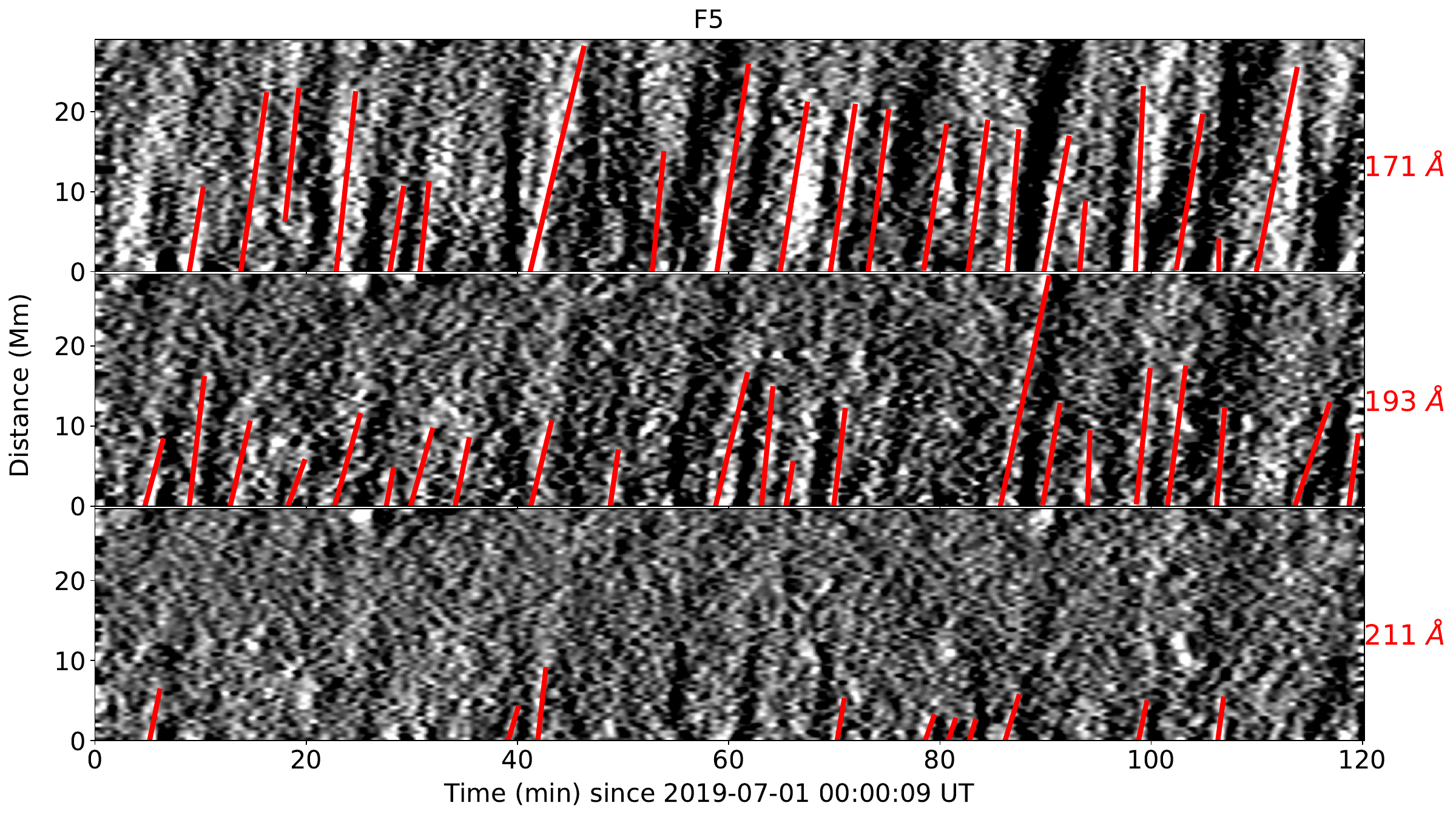}
    \rm{I}. The peak of the plume.
    \includegraphics[width=\linewidth]{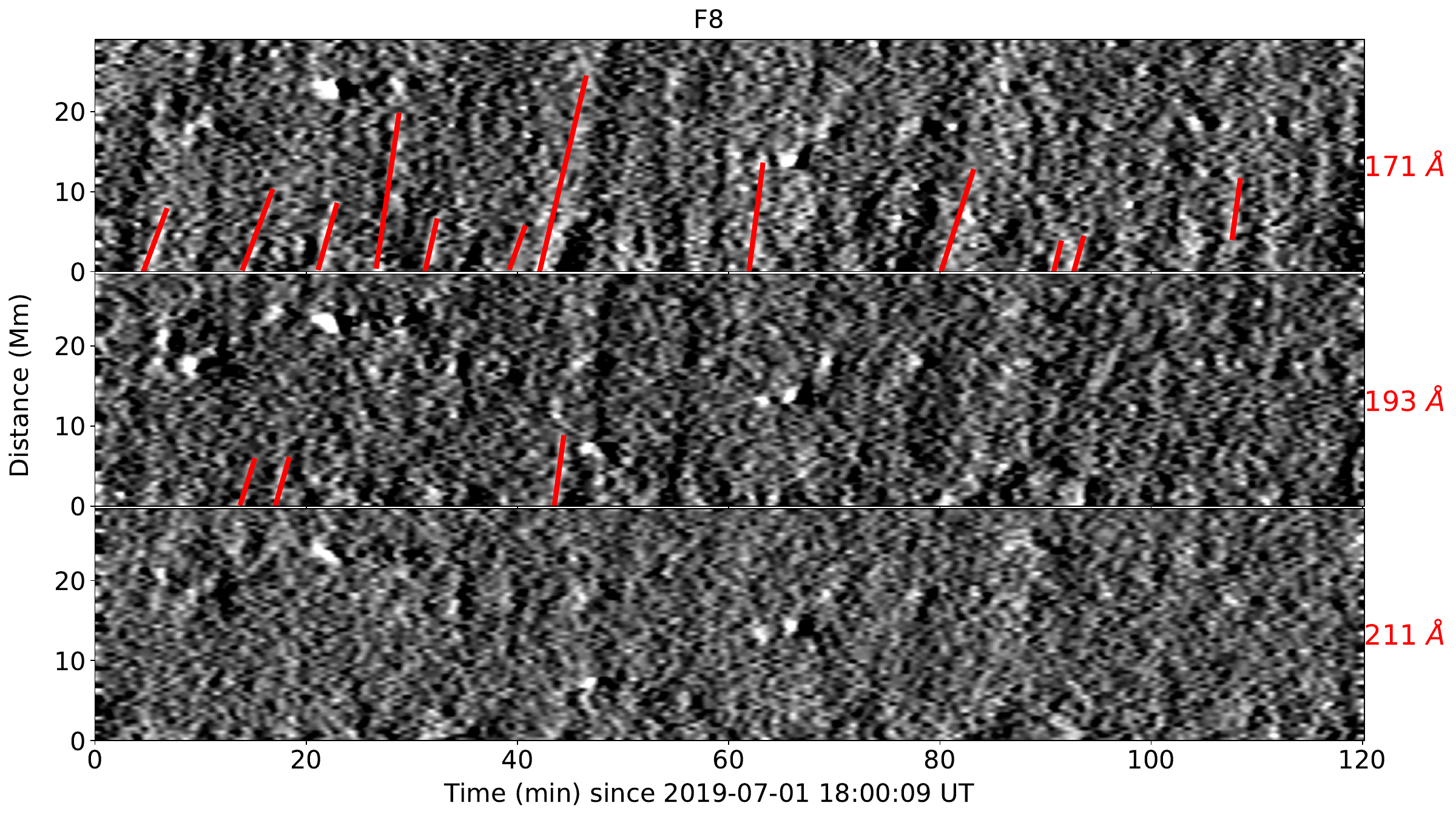}
    \rm{II}. End of the plume.
    \caption{Panel \rm{I}: Space-time plot during the peak brightening. Panel \rm{II}: Space-time plot at the end of the plume. From top to bottom for each panel, we display the plot in 171~{\AA},193~{\AA}, and 211~{\AA} channels. Red lines are drawn to show outflow lines.}
    \label{fig:xt_plot34}
\end{figure}

\begin{figure}[htpb!]
    \centering
    \includegraphics[width=\linewidth]{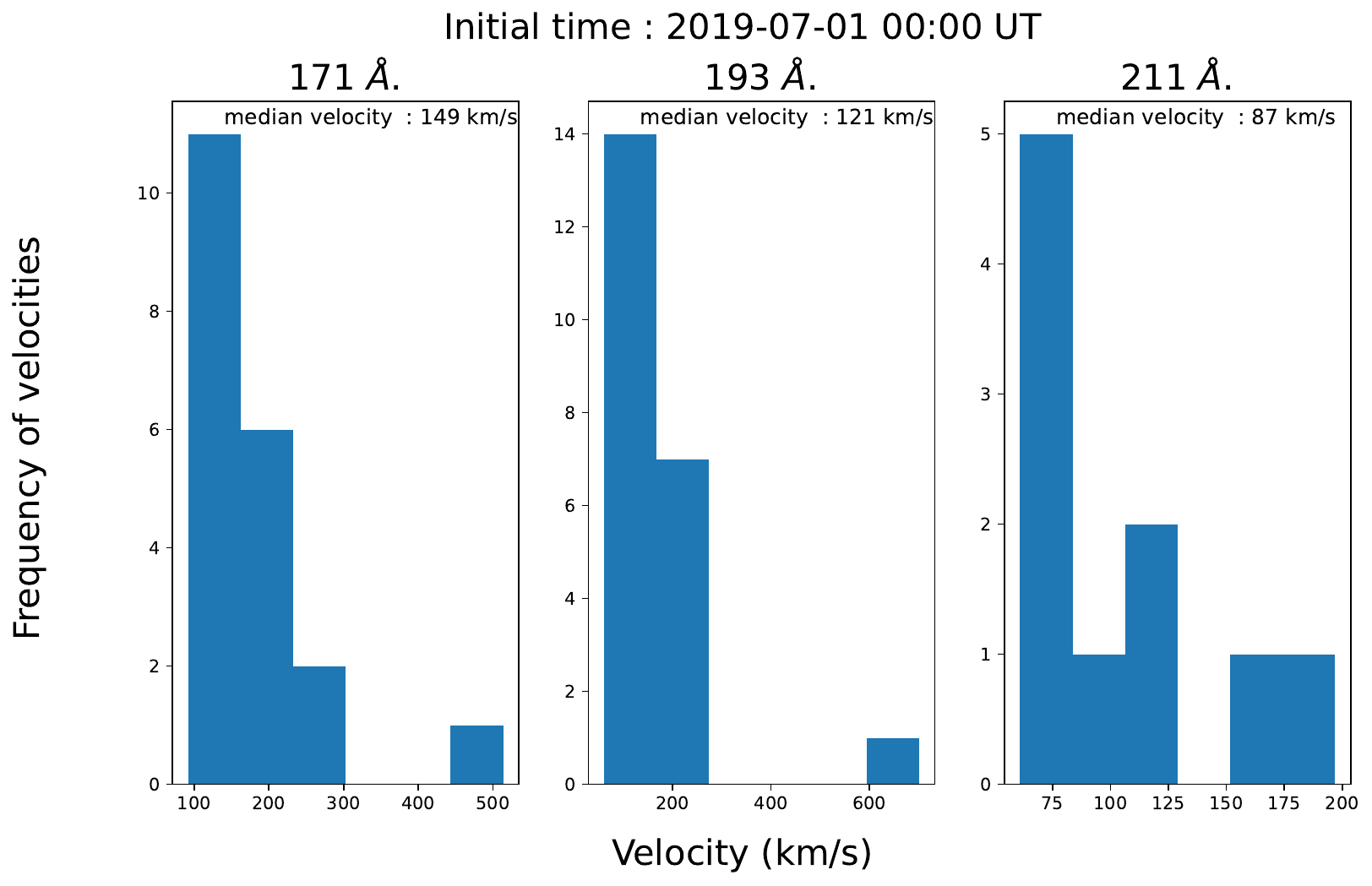}
    \caption{Histogram obtained for the xt-plot F5. Histograms of the plane of sky velocity were obtained using space-time plots for 171~{\AA} (left), 193 {\AA} (middle), and 211~{\AA} (right) panels.}
    \label{fig:velocity_hist}
\end{figure}

Using the above-described space-time maps, we can derive the plasma flow speed in the plane of the sky by qualitatively fitting straight lines to the ridges as shown in Figs.~\ref{fig:xt_plot12}~\&~\ref{fig:xt_plot34}. We compute the median velocity if we find more than six discernible outflows (straight lines in Fig~\ref{fig:xt_plot12} and Fig~\ref{fig:xt_plot34}). In Fig~\ref{fig:velocity_hist}, we show the histogram of velocities obtained for F5, while in Table~\ref{tab:velocity_table}, we show the median velocities for 3 EUV passbands. We obtain the speeds of all the ridges, and the speeds range from 100{--}150~{\kms} in 171~{\AA},  100{--}120~{\kms} in 193~{\AA}, 50{--}100~{\kms} in 211~{\AA} observations.

\begin{deluxetable}{ccccc}[h!]
\tabletypesize{\footnotesize}
\tablewidth{0pt}
\tablecaption{Velocities obtained from xt diagram \label{tab:velocity_table}}
\tablehead{
\colhead{Start time}   &   \colhead{171{\AA}}   &   \colhead{193{\AA}}   &   \colhead{211{\AA}} & \colhead{}\\
\colhead{UT}   &   \colhead{km/s}   &   \colhead{km/s}   &   \colhead{km/s} & \colhead{}
}
\startdata
*2019-06-30 00:00   &   125   &   211   &   -   & Just formed\\
*2019-06-30 06:00  &   87   &   -   &   -   & First dip in brightness\\
2019-06-30 12:00   &   105   &   -   &   -  & -\\
2019-06-30 18:00   &   111   &   102   &   64   &-\\
*2019-07-01 00:00   &   149   &   121   &   87  & Peak brightness\\
2019-07-01 06:00   &   143   &   115   &   -&   -\\
2019-07-01 12:00   &   135   &   100   &   101& -\\
*2019-07-01 18:00   &   83   &   -   &   -  & Disappearance phase\\
\enddata
\tablecomments{Space-time plots for * marked shown in the main paper and rest in the Appendix.}
\end{deluxetable}

\subsection{Thermal Structure}\label{DEM Structure}
Having studied the kinematic properties at various instants of time, we now study the thermal structure of the plume and derive temperature as a function of altitude. For this purpose, we derive the DEM from the six EUV intensities of AIA. The data were normalized by exposure time and averaged over 5 minutes for each of the six bands. After doing that, we use $aia\_prep.pro$ routine, provided with SolarSoft of IDL \citep{sswidl} to correct for roll angle and plate scale. We analyzed the thermal structure from 2019-06-30 00:00 UT to 2019-07-02 00:00 UT, when the plume was at the peak of its intensity, with an effective cadence of 1 hour.

\begin{figure}[htpb!]
    \centering
    \includegraphics[scale=0.7]{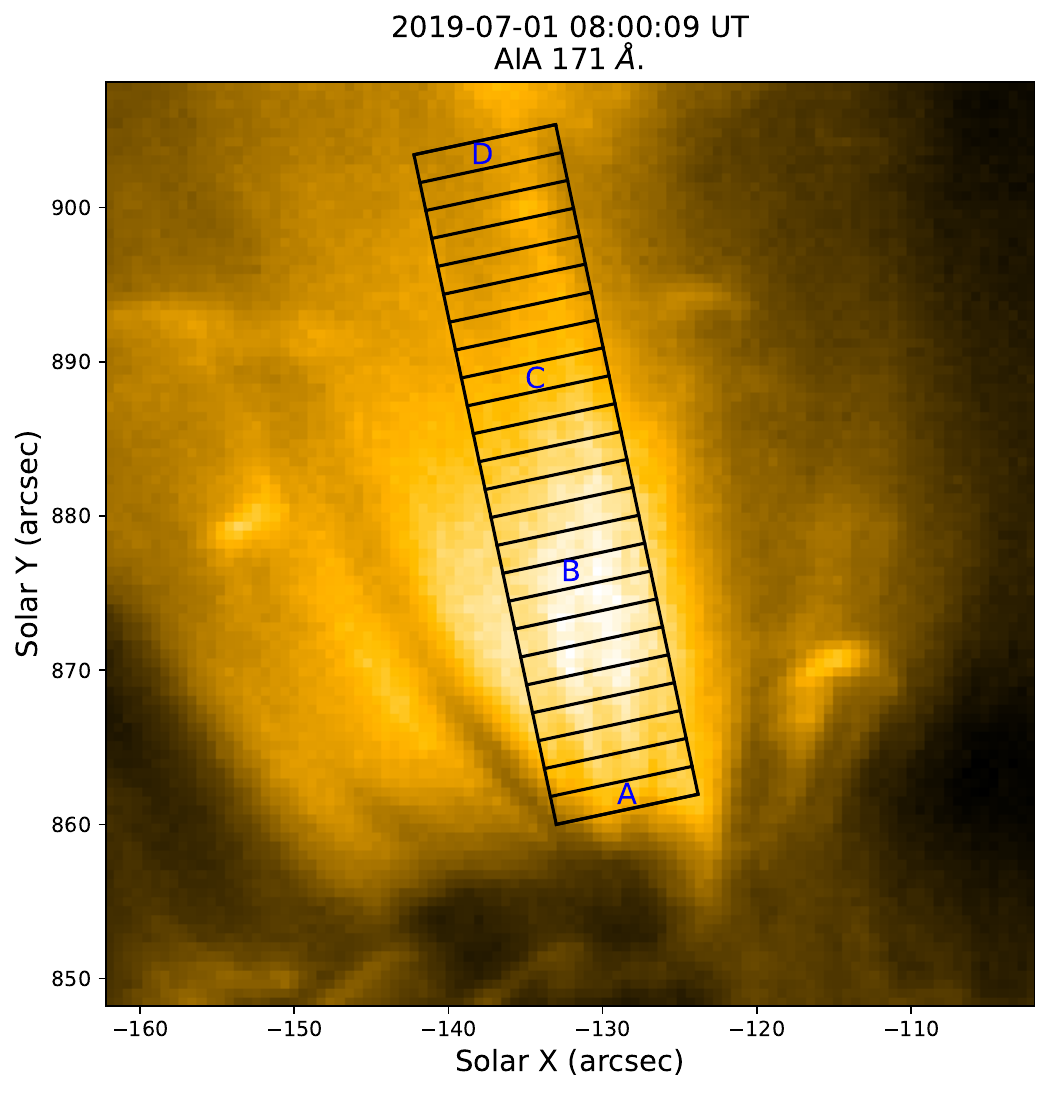}
    \caption{AIA 171~{\AA} image overplotted the region that is considered for DEM analysis. A, B, C, and D are the regions for which the DEM curve has been shown in Fig.~\ref{fig:dem_curve}.}
    \label{fig:dem_slit}
\end{figure}

Note that we do not construct pixel-to-pixel DEM. Instead, we place a slit with width $10\arcsec$ and height $45\arcsec$ along the plume length and slice the slit into 24 equally spaced regions along the length of the slit, as shown in Fig.~\ref{fig:dem_slit}. We obtain the average intensity for each region for all the six EUV passbands of AIA, which is then used to obtain DEM. Moreover, we use Monte Carlo analysis to estimate the DEM uncertainties. Photon counting on a CCD has a Poissonian error of $\sqrt{n}$, where $n$ is the intensity in digital number. So, for every region on the artificial slit, we can construct several intensities with error $\sqrt{n}$ by drawing samples from a Gaussian distribution with mean $n$ and standard deviation of $\sqrt{n}$~\footnote{Poisson distribution can be approximated with Gaussian distribution for large mean value.}. Hence, we generate multiple DEM realizations within error in intensity for each block on the slit. We use the regularized DEM inversion method by \cite{hannah-kontar1, hannah-kontar2} and obtained DEM for all the blocks of the slit mentioned above. We consider a temperature range from $\mathrm{\log T/[K]}$ = 5.5 to 6.5, with a spacing of $\Delta\mathrm{\log T/[K]}$ = 0.01.

\begin{figure}[htpb!]
    \centering
    \includegraphics[scale=0.5]{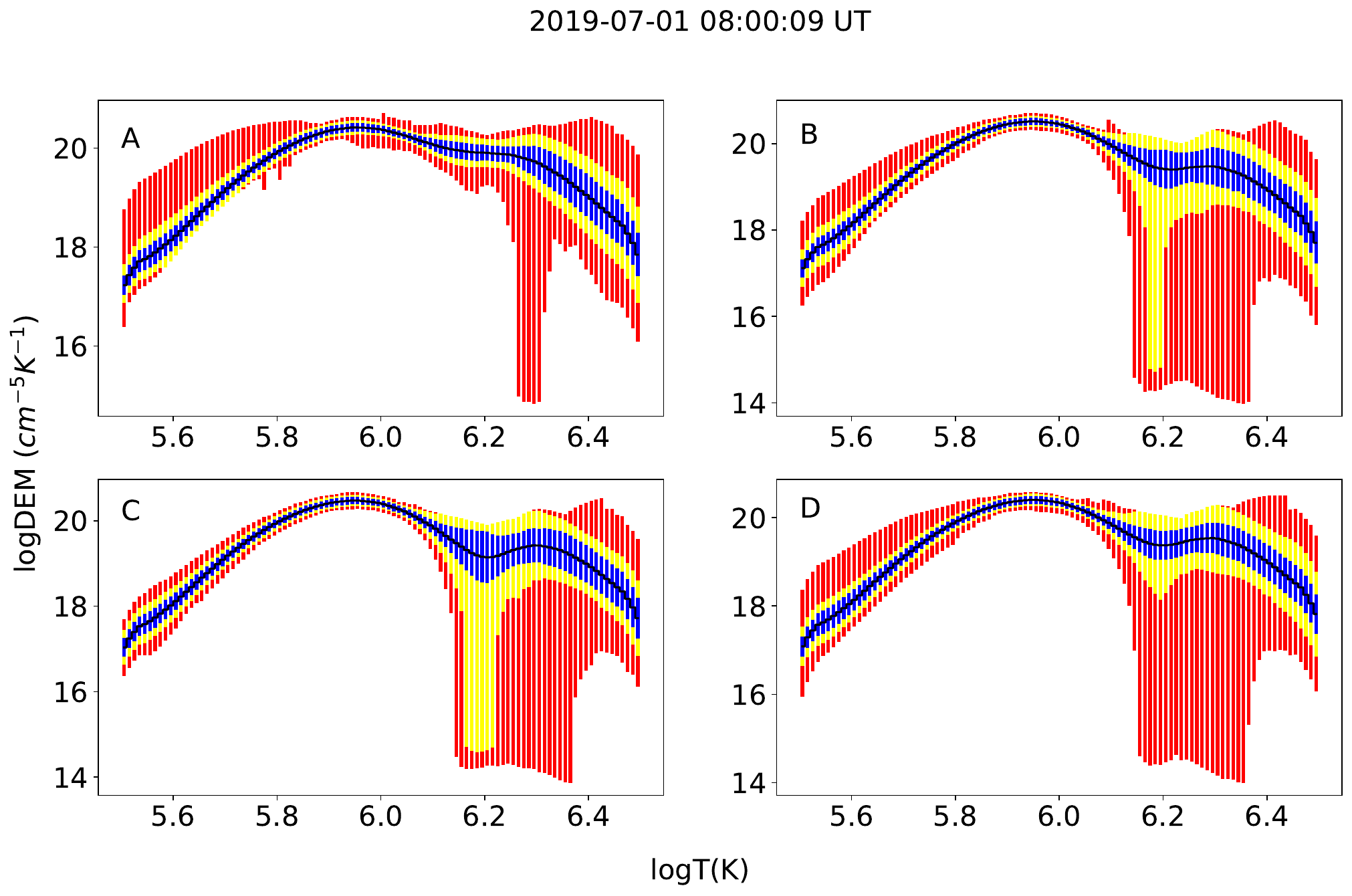}
    \caption{DEM curves for the four blocks marked by (A, B, C, D) in the Fig.~\ref{fig:dem_slit}. 50$\%$ Monte Carlo solutions are within the blue region, 80$\%$ in the yellow region, and 95$\%$ in the red region. The temperature range has been taken between $logT = 5.5$ to $logT = 6.5$.}
    \label{fig:dem_curve}
\end{figure}

In Fig.~\ref{fig:dem_curve}, we show four such DEM curves from 4 blocks (marked by A, B, C, D) of Fig.~\ref{fig:dem_slit}. In the plot, 50$\%$ of the Monte Carlo realizations are within blue bars, 80 $\%$ within yellow, and 95 $\%$ are within red bars. The DEM curve peaks around $\log\,T(K) \sim$ 5.9. However, significant emissions also can be observed up to $\log\,$T(K) $\sim$ 6.2.
We also compute the DEM weighted temperature, defined as: 
\begin{equation}
    T_e = \frac{\int_{T_1}^{T_2} T~\mathrm{DEM(T)}\, dT }{\int_{T_1}^{T_2} \mathrm{DEM(T)}\, dT }.
\end{equation}

\begin{figure}[htpb!]
    \centering
    \includegraphics[width=\linewidth]{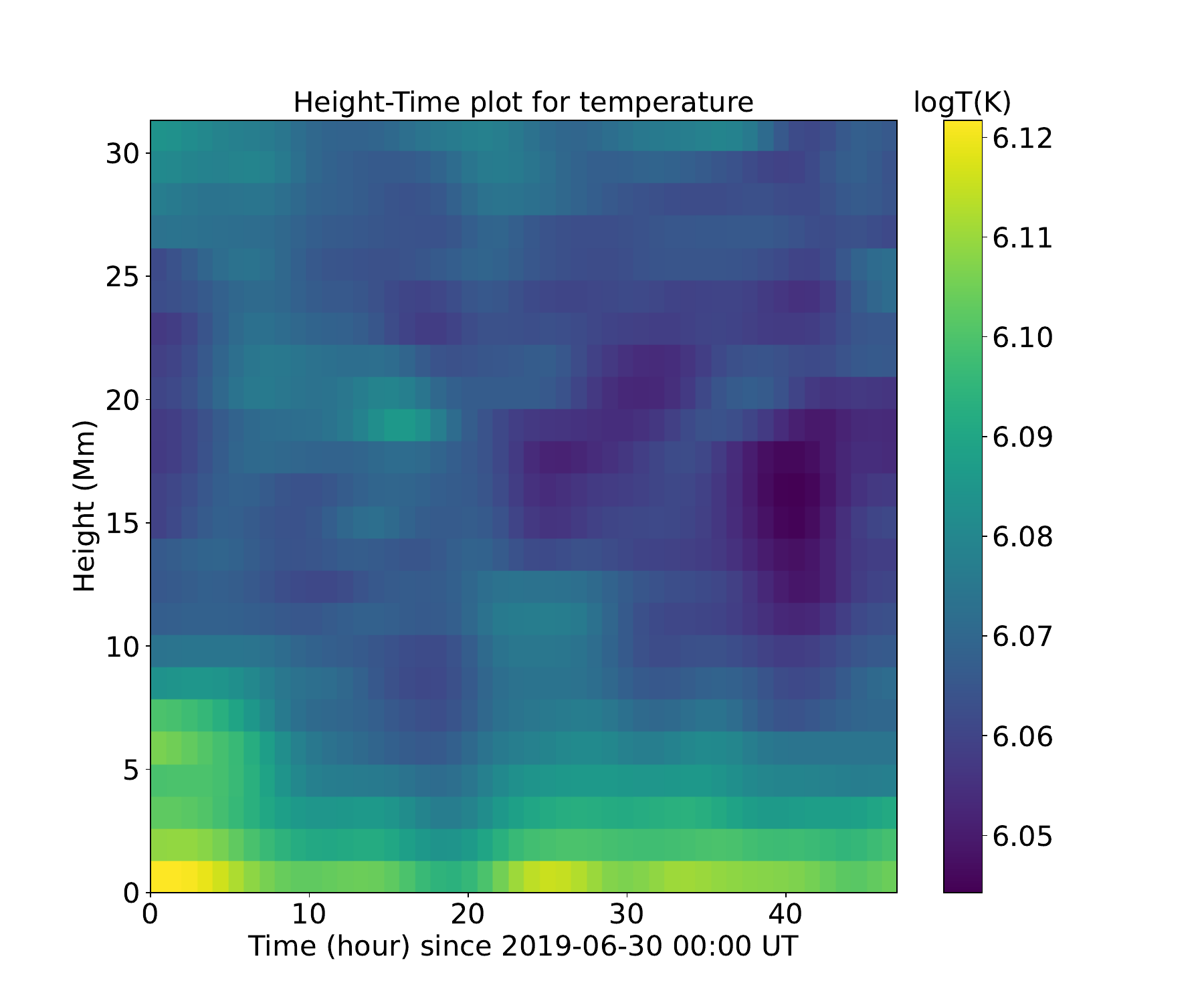}
    \caption{Variation of the DEM weighted temperature along the length of the plume as a function of time. Time starts from 2019-06-30 00:00 UT to 2019-07-02 00:00 UT.}
    \label{fig:temp_evolution}
\end{figure}

\begin{figure}[H]
    \centering
    \includegraphics[width=\linewidth]{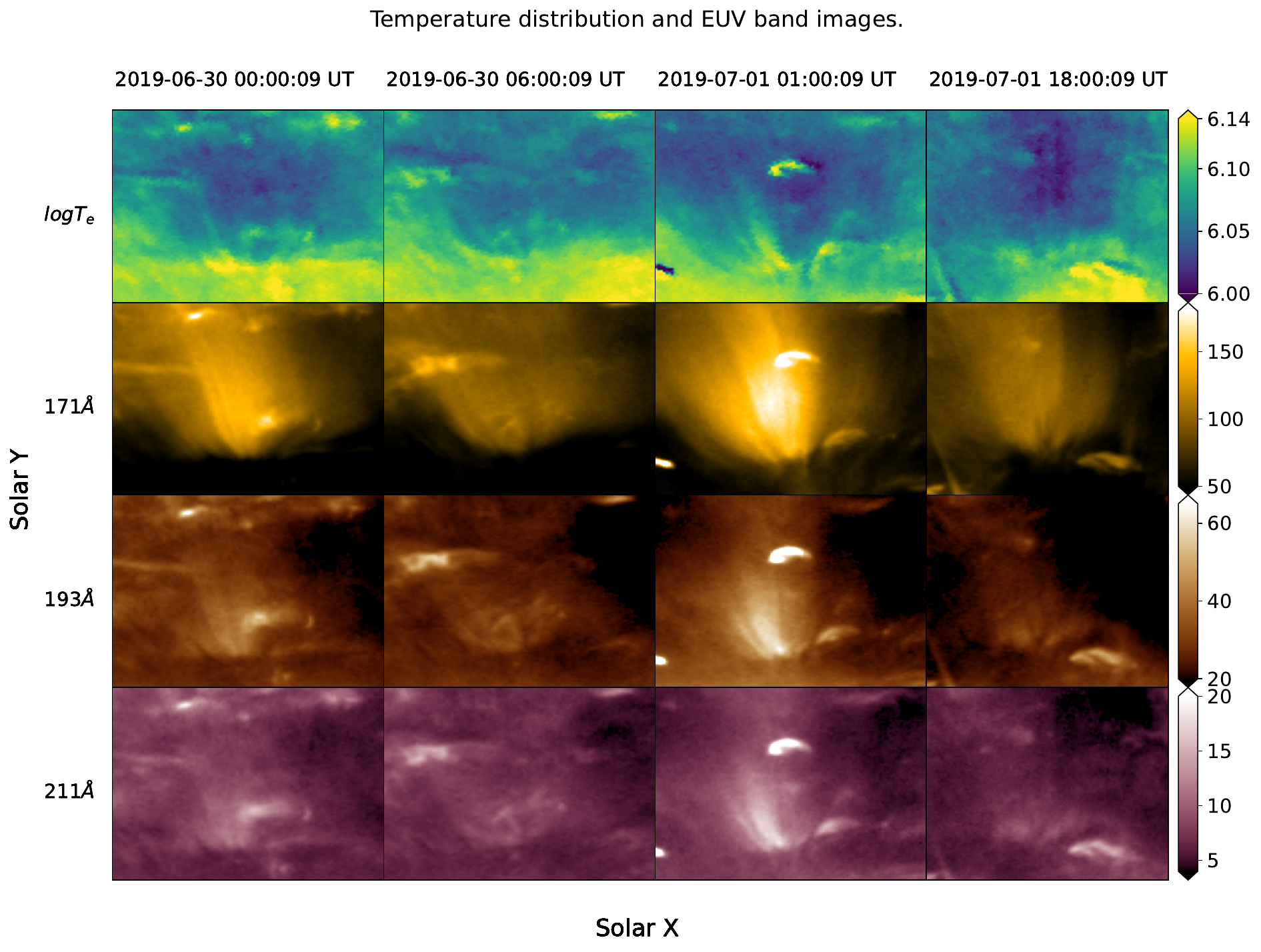}
    \caption{Comparison of the DEM weighted temperature map (top row) with the images obtained in three AIA passbands, namely 171~{\AA} (2nd row), 193~{\AA} (third row), and 211~{\AA} (bottom row). The leftmost column represents the time when the plume just formed, then the first dip in the plume brightness, then the peak of the plume, and finally, the disappearance of the plume.}
    \label{fig:AIA_Te}
\end{figure}

We obtain temperature along the plume length for two days with an effective cadence of 1 hour and plot temperature variation along length and time as shown in Fig.~\ref{fig:temp_evolution}. The plot reveals that the plume is predominantly isothermal with $\mathrm{\log T/[K]}=6.08$. However, the temperature from the base to about $\sim$5~Mm may have temperatures $\mathrm{\log T/[K]}\approx6.12$. We also note that the body of the plume, at $\approx10-25$ Mm, may exhibit temperatures lower than $\mathrm{\log T/[K]}\approx6.05$
We further obtained DEM-weighted temperature for each pixel and created temperature maps. In Fig.~\ref{fig:AIA_Te}, we show DEM-weighted temperature maps (top row) and images in three EUV channels, namely, 171~{\AA}, 193~{\AA} and 211~{\AA} in the bottom three rows, respectively. 

The first column in Fig.~\ref{fig:AIA_Te} corresponds to the time when the plume starts to form. The faint signal from the plume is seen in AIA 193~{\AA} and 211~{\AA} images, but it is seen clearly in the 171~{\AA} channel. Therefore, in the temperature map, hardly any structures are above $\mathrm{\log T/[K]}\sim6.14$. Images in AIA 193~{\AA} and AIA 211~{\AA} are even fainter during the first dip ( second column), showing no temperature structures above $\mathrm{\log T/[K]}\sim6.14$. In the third column, when the plume is at the peak of its brightness, we can see clear structures in all three passbands and a temperature structure inside the plume above $\mathrm{\log T/[K]}\sim6.14$. Hence, the plume structure is not isothermal but has temperature variations inside the plume. Finally, in the last column, we find a significant decrease in the plume brightness with a very faint temperature structure.

\section{Summary and Discussions}\label{summary}
In this paper, we present a detailed study of a solar coronal plume in a CH. The plume peaked on 2019-07-01, and we study it through its birth, evolution, and eventual disappearance over several days. The plume begins with several jets and jetlet events at the base, resulting in a gradual development of the plume haze. The haze development occurs over $\approx46$ hours, while the plume then forms very rapidly within a period of $\approx6$ hours. The plume base remains highly dynamic through the life cycle, with multiple episodes of jets and jetlets. This observation aligns with the observation of jets and transient activity being precursors of plumes~\citep{Raouafi-2009,raouafi-stenborg-2014}. Although plumes are often observed above bright points \citep{wang-muglach-2008,Pucci-2014}, we do not see any conspicuous signature of bright points at the plume's base.

We then analyze the light curves for the plume from the start to the end of the plume for all the SDO/AIA EUV bands. We find that all the light curves have similar profiles except the 304~{\AA}. The similarity of spectral response in these EUV passbands can be explained through the multi-thermal nature of the plume plasma. We find the plume to be the brightest in AIA 171~{\AA} passband, which may be inferred to have a dominant contribution from \ion{Fe}{9} 171~{\AA} (at $\mathrm{\log T/[K]} \approx 5.85$)~\citep{O'Dwyer-2010,Boerner-2012}. Similarly, the strongest contributions in 193~{\AA} would come from \ion{Fe}{12} 193.51~{\AA} (at $\mathrm{\log T/[K]} \approx 6.2$), while in 211~{\AA} would be from \ion{Fe}{14}~211.32~{\AA} (at $\mathrm{\log T/[K]} \approx 6.3$). We also note that ions like \ion{Fe}{10} (at $\mathrm{\log T/[K]} \approx 6.05$) have several transitions contributing to emission in AIA 94~{\AA}, 171~{\AA}, 193~{\AA}, and 335~{\AA}, which would explain similar intensity profiles in these different passbands. Since AIA 304~{\AA} is dominated by the chromospheric {\rm He II} line forming in optically thick conditions, we do not see qualitatively similar light curve structure in this passband. However, we observe signatures of outflows at the base of the plume. Such outflows also have been observed by \citep{cho}.

We study the evolution of the positive ($\Sigma B_+$) and negative ($\Sigma B_-$) magnetic flux at the base and compare its evolution with the light curve of the plume in  Fig.~\ref{fig:171A_B} and ~\ref{fig:correlation}. At initial times, we find the plume intensity independent of the $\Sigma B_+$ and $\Sigma B_-$. During the main phase of the plume, an increase in $\Sigma B_+$ is correlated with an increase in AIA 171~{\AA} intensity. With $\Sigma B_-$, we, however, find that an initial increase does not correlate with an increase in plume intensity. We see the opposite -- $\Sigma B_-$ reduced, and a correlated increase in plume intensity is seen. $\Sigma B_-$ shows multiple episodes of rise and fall of flux while the plume intensity keeps increasing. Towards the end of the main phase, the intensity remains constant and becomes insensitive to changes in the dominant polarity, $\Sigma B_+$. The plume intensity drops and remains constant towards the end of the main phase, while $\Sigma B_-$ approximately goes back to the pre-jet period values. 

The dispersion of dominant polarity at the end of the plume is similar to the observations reported by \cite{raouafi-stenborg-2014}. Correlations analysis has been done for plume intensity in literature. For instance,~\cite{uritsky-2021} study the relation between the number of detectable jetlets and plume brightness and find a positive correlation between them. Contrarily, \cite{Pucci-2014} do not find any correlations between the plume and bright point intensity variations. While the magnetic field may have some relation with bright points and the statistics of jetlets, we demonstrate a causal relationship between the photospheric magnetic field and the plume brightness.

We measure flow velocities in the plume by constructing space-time plots (xt-plots) of running difference images. We find velocities of $\approx100-150$ km/s in 171~{\AA}, $\approx100-120$ km/s in 193~{\AA}, and $\approx50-100$ km/s in 211~{\AA}. \cite{tian-2011} report speeds of $\sim $ 120 $\pm$ 30 km/s in 171 , 193 and 211 {\AA} bands while \cite{Pucci-2014} reported median speeds of ~185 km/s for AIA 171 {\AA} channel, and 143 km/s and 124 km/s for 193 {\AA} and 211 {\AA} passband. \cite{krishan_prashad-2011} report velocities of 100 -- 120 km/s in the 171 {\AA}, and 130 -- 140 km/s in 193 {\AA} passband. We note that the measured flow speeds in this work align with those of previous studies. If we consider the sound speed $c_s = 166 T^{1/2}$ km/s \citep{priest-2014} and characteristic temperatures ($\mathrm{\log T/[K]}$) in the 171 {\AA}, 193 {\AA} and 211 {\AA} bands of 5.8 , 6.2 and 6.3 respectively \citep{AIA1}, we would obtain $c_s \sim$ 130 km/s, 210 km/s, 230 km/s respectively. Due to a lack of dependence of flow speeds on different temperatures (passbands), \cite{tian-2011,Pucci-2014} interpret that the flow speeds are propagating disturbances rather than waves.~\cite{krishan_prashad-2011}, on the other hand, interpret the presence of slow magneto-acoustic waves as they report higher speeds for higher temperature bands, although they have uncertainties of $\approx$20$\%$ that would lead to different interpretations. We find that the higher temperature bands have lower median speeds, as did \cite{Pucci-2014}. Hence, we are inclined to interpret the outflows detected here as propagating disturbances rather than waves. However, we find several instances of speeds comparable to the sound speed (see Fig.~\ref{fig:velocity_hist}), so we do not rule out the presence of any waves. We note that it is plausible to conclude that plume hosts both waves and propagating disturbances \citep{de-forest-1998,krishnaprasad-2012,Pucci-2014}, dominated by propagating disturbances, later in larger numbers.

We find many more instances of outflows during the peak of the plume's brightness rather than during the dimming of the plume brightness (see Fig.~\ref{fig:xt_plot12}~\&~\ref{fig:xt_plot34}). Thus, the higher number of outflows occurring at the plume's base may feed plasma to the plume, raising its brightness. This contrasts with the \cite{Pucci-2014}, who find the number of outflows to be independent of the phase of the plume. However, we note that \cite{uritsky-2021} finds a linear correlation between the number of detectable edges inside the plume and the plume brightness. So, these detectable edges might be the flows we detect here.

We further note that although we find observable signatures of outflows in 171~{\AA} channel very clearly, most of the time we do not find signatures of outflows in the 211~{\AA} channel. This may be attributed to the likely presence of high shot noise in the 211~{\AA} channel, as has been suggested by \citep{uritsky-2013}. More advanced techniques like surfing transformation \citep{uritsky-2013} may provide velocity signatures in 211~{\AA} channel.

We finally study the temperature structure of the plume through DEM analysis~\citep{hannah-kontar1,hannah-kontar2}. After reducing the spatial resolution of the plume as given in Fig.~\ref{fig:dem_slit}, we obtain the DEM weighted temperature along the length of the plume as a function of time. Over larger spatial scales, we find the plume body to be nearly isothermal, similar to results in existing literature  \citep[see, e.g.,][]{wilhelm}. However, high-resolution observation suggests that the plume is substantially multi-thermal, containing high-temperature outflows along those high-temperature structures. Observation of such internal structure within plumes is again consonant with existing literature \citep{raouafi-stenborg-2014,uritsky-2021}.
Regarding the morphology and lifetime of the plume, it is evident that magnetic field concentrations at the base of the plume play an important role in the formation and maintenance of the plumes but is not well studied how the magnetic field evolves with the evolution of the plume or vice versa. There are other larger scale structures like pseudo streamer (\cite{guhathakurata-1995},\cite{wang-2007},\cite{gannouni-2024}) that resemble in morphology but in larger scale and time. We note that such structures are possibly self-similar but with differences in the strength and location of the magnetic field, and transients happening at the base. To make firm conclusions on this aspect, it is required to perform further dedicated study, which is beyond the scope of this paper.

\section{Conclusions}\label{conclusions}
\begin{figure}[htpb!]
    \centering
    \includegraphics[width=\linewidth]{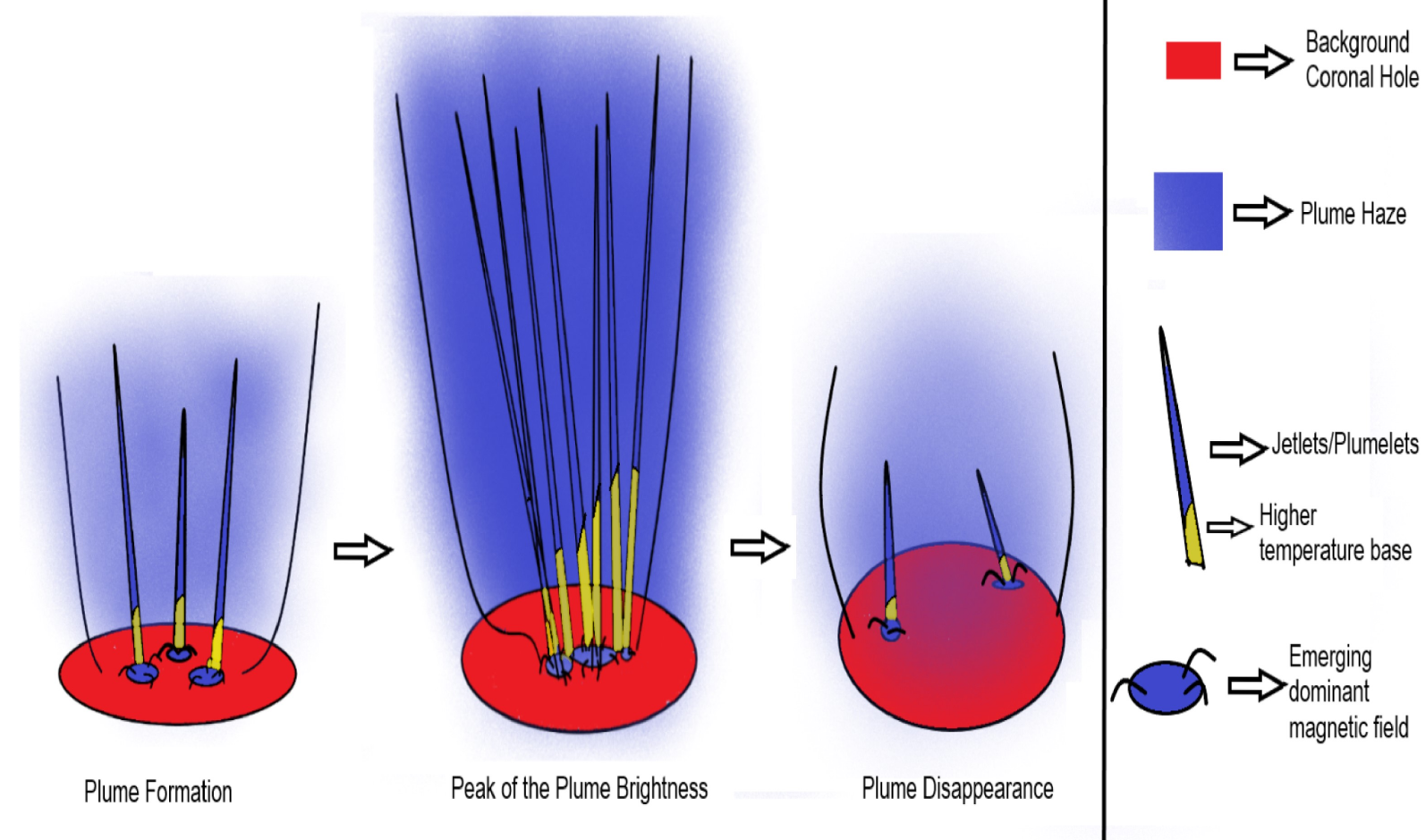}
    \caption{A cartoon showing the main phases of the plume. During the formation of the plume, we have less number of jets and jetlets activities going on, while at the peak of the intensity we have a higher number of jetlets activities that is higher number of outflow activities going on, and at the end of the plume the dominant base magnetic field fragments and disperse away leading to the plume disappearance.}
    \label{fig:plume_cartoon}
\end{figure}

Our observations suggest that the dynamics of underlying magnetic flux largely drive the dynamics of the plume. The initial jets and jetlets observed at the plume's base support the hypothesis that magnetic reconnection between emerging magnetic bipoles and pre-existing open field lines plays a vital role in plume formation, as also proposed by earlier models and observations \citep{wang-1998,raouafi-stenborg-2014,Raouafi-2009,uritsky-2021,Pucci-2014}. The strong correlation between the emerging magnetic field and the plume's brightness (Fig.~\ref{fig:correlation}) suggests a strong causal association between the magnetic flux at the base and the plume intensity. Specifically, the loss of the minority flux is crucial for sustaining the plume's existence and determining its intensity.

The higher number of outflows and observations of higher temperature internal structures at the base of the plume during its peak intensity suggest jetlet activities by impulsive reconnection. Such fine structures may manifest into solar wind structures at heliocentric distances. In Fig.~\ref{fig:plume_cartoon}, we summarize the physical picture suggested by our observations. During the formation of the plume, we see plume haze gradually appearing which ultimately leads to peak brightness with a higher number of internal structures with higher temperatures at the base of the plume. The disappearance of the plume occurs with the fragmentation and dispersal of the magnetic field concentrations at the base of the plume.

This puts further constraints on future modelling of plumes that should take into account the internal thermal structures (jets and jetlets) and should be able to explain the intensity variation against the magnetic field at the base of the plume.


\section*{}
We thank the referee for the constructive feedback on the manuscript. We acknowledge the CEFIPRA grant 6904-2. U.V. gratefully acknowledges support by NASA contracts NNG09FA40C (IRIS) and 80GSFC21C0011 (MUSE). B.M. acknowledges valuable discussions and suggestions with Prof. Helen Mason, and Nived V N. U.V. acknowledges discussions with Ron Moore at the AGU meeting in 2023. We acknowledge the use of data from AIA and HMI. AIA and HMI are instruments on board SDO, a mission for NASA's Living With a Star program.

\appendix

\begin{deluxetable}{cc|cc|cc}[ht!]
\tabletypesize{\footnotesize}
\tablewidth{0pt}
\tablecaption{Plume base positions. The position corresponds to the maximum magnetic field strength beneath the plume base.} \label{tab:plume_base1}
\tablehead{
\colhead{Date}   &   \colhead{Position}   &   \colhead{Date}   &   \colhead{Position}   &   \colhead{Date}   &   \colhead{Position}\\
\colhead{UT}   &   \colhead{(\arcsec,\arcsec)}   &   \colhead{UT}   &   \colhead{(\arcsec,\arcsec)}   &   \colhead{UT}   &   \colhead{(\arcsec,\arcsec)}
}
\startdata
2019-06-27 23:59   &   [-277,874]   &   2019-06-29 20:59   &   [-197,866]   &   2019-07-01 17:59   &   [-97,855]\\
2019-06-28 02:29   &   [-271,876]   &   2019-06-29 23:29   &   [-192,865]   &   2019-07-01 20:29   &   [-90,855]\\
2019-06-28 04:59   &   [-270,875]   &   2019-06-30 01:59   &   [-187,864]   &   2019-07-01 22:59   &   [-84,855]\\
2019-06-28 07:29   &   [-260,875]   &   2019-06-30 04:29   &   [-182,863]   &   2019-07-02 01:29   &   [-75,854]\\
2019-06-28 09:59   &   [-249,877]   &   2019-06-30 06:59   &   [-176,863]   &   2019-07-02 03:59   &   [-71,856]\\
2019-06-28 12:29   &   [-255,873]   &   2019-06-30 09:29   &   [-173,862]   &   2019-07-02 06:29   &   [-67,855]\\
2019-06-28 14:59   &   [-250,874]   &   2019-06-30 11:59   &   [-168,861]   &   2019-07-02 08:59   &   [-63,857]\\
2019-06-28 17:29   &   [-246,874]   &   2019-06-30 14:29   &   [-163,860]   &   2019-07-02 11:29   &   [-58,856]\\
2019-06-28 19:59   &   [-239,873]   &   2019-06-30 16:59   &   [-156,861]   &   2019-07-02 13:59   &   [-53,857]\\
2019-06-28 22:29   &   [-237,872]   &   2019-06-30 19:29   &   [-153,859]   &   2019-07-02 16:29   &   [-51,858]\\
2019-06-29 00:59   &   [-234,871]   &   2019-06-30 21:59   &   [-149,858]   &   2019-07-02 18:59   &   [-41,857]\\
2019-06-29 03:29   &   [-227,869]   &   2019-07-01 00:29   &   [-145,858]   &   2019-07-02 21:29   &   [-38,857]\\
2019-06-29 05:59   &   [-227,869]   &   2019-07-01 02:59   &   [-140,857]   &   2019-07-02 23:59   &   [-32,859]\\
2019-06-29 08:29   &   [-220,868]   &   2019-07-01 05:29   &   [-131,857]   &   2019-07-03 02:29   &   [-27,857]\\
2019-06-29 10:59   &   [-215,868]   &   2019-07-01 07:59   &   [-124,856]   &   2019-07-03 04:59   &   [-20,857]\\
2019-06-29 13:29   &   [-211,867]   &   2019-07-01 10:29   &   [-114,856]   &   2019-07-03 07:29   &   [-13,854]\\
2019-06-29 15:59   &   [-206,867]   &   2019-07-01 12:59   &   [-109,855]   &   2019-07-03 09:59   &   [-8,857]\\
2019-06-29 18:29   &   [-201,866]   &   2019-07-01 15:29   &   [-103,854]   &   2019-07-03 12:29   &   [-2,860]\\
\enddata
\end{deluxetable}

\begin{deluxetable}{ccc}[ht!]
\tabletypesize{\footnotesize}
\tablewidth{0pt}
\tablecaption{Slit information for space-time plot. Angle is with respect to North and toward East. Slit is 5{\arcsec} wide and 40{\arcsec} long. \label{tab:slit_info_xt_plot}}
\tablehead{
\colhead{Time}   &   \colhead{Bottom left position}   &   \colhead{Angle}\\
\colhead{UT}   &   \colhead{[arcsec,arcsec]}   &   \colhead{Degree}
}
\startdata
2019-06-30 00:00   &   [-195, 873]   &   -10\\
2019-06-30 06:00   &   [-182, 874]   &   -18\\
2019-06-30 12:00   &   [-169, 870]   &   -7\\
2019-06-30 18:00   &   [-165, 866]   &   -15\\
2019-07-01 00:00   &   [-151, 865]   &   -10\\
2019-07-01 06:00   &   [-139, 865]   &   -18\\
2019-07-01 12:00   &   [-121, 865]   &   -12\\
2019-07-01 18:00   &   [-105, 865]   &   -8\\
\enddata
\end{deluxetable}

\begin{figure}[H]
    \centering
    \includegraphics[width=\linewidth]{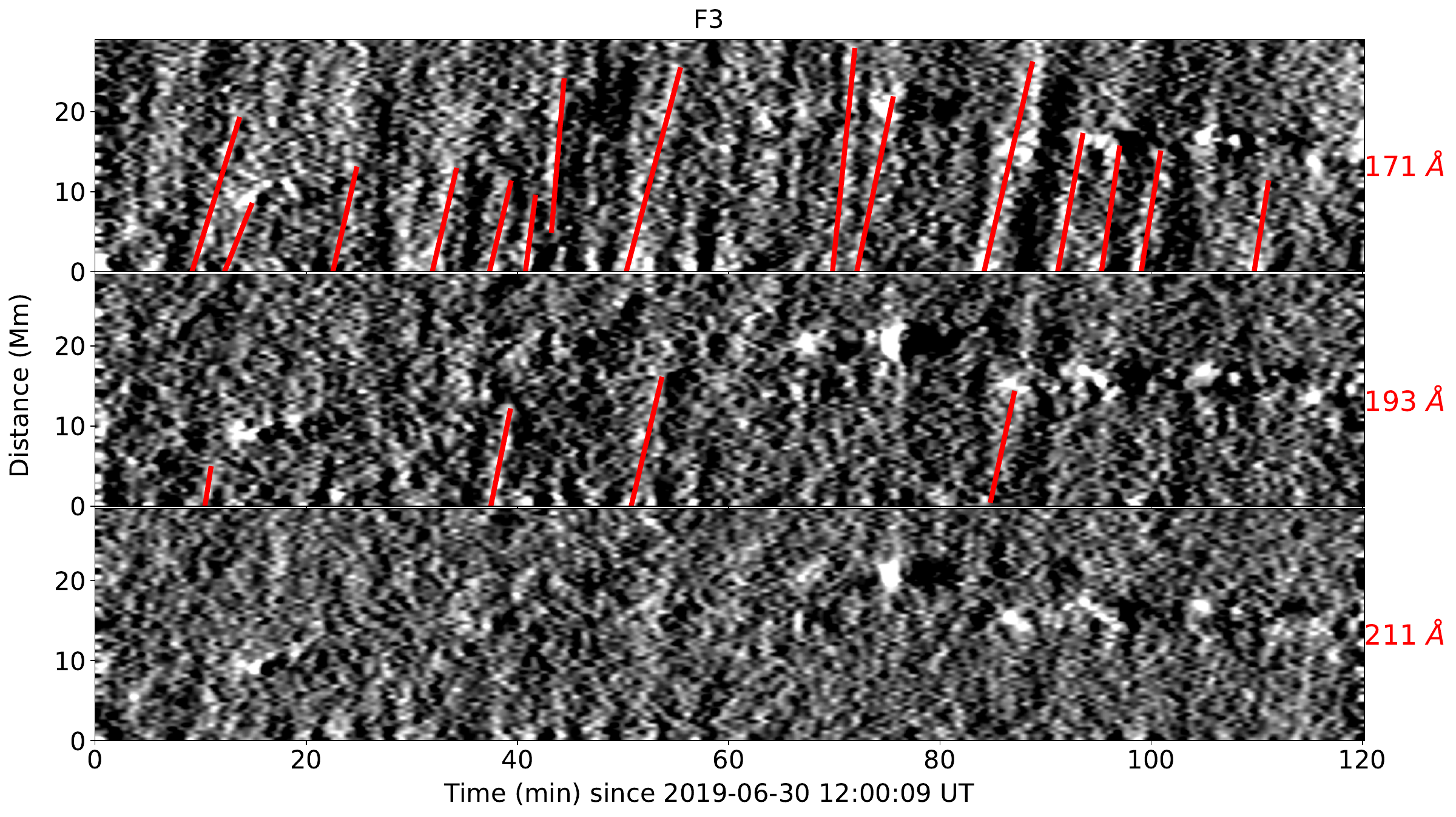}

    \includegraphics[width=\linewidth]{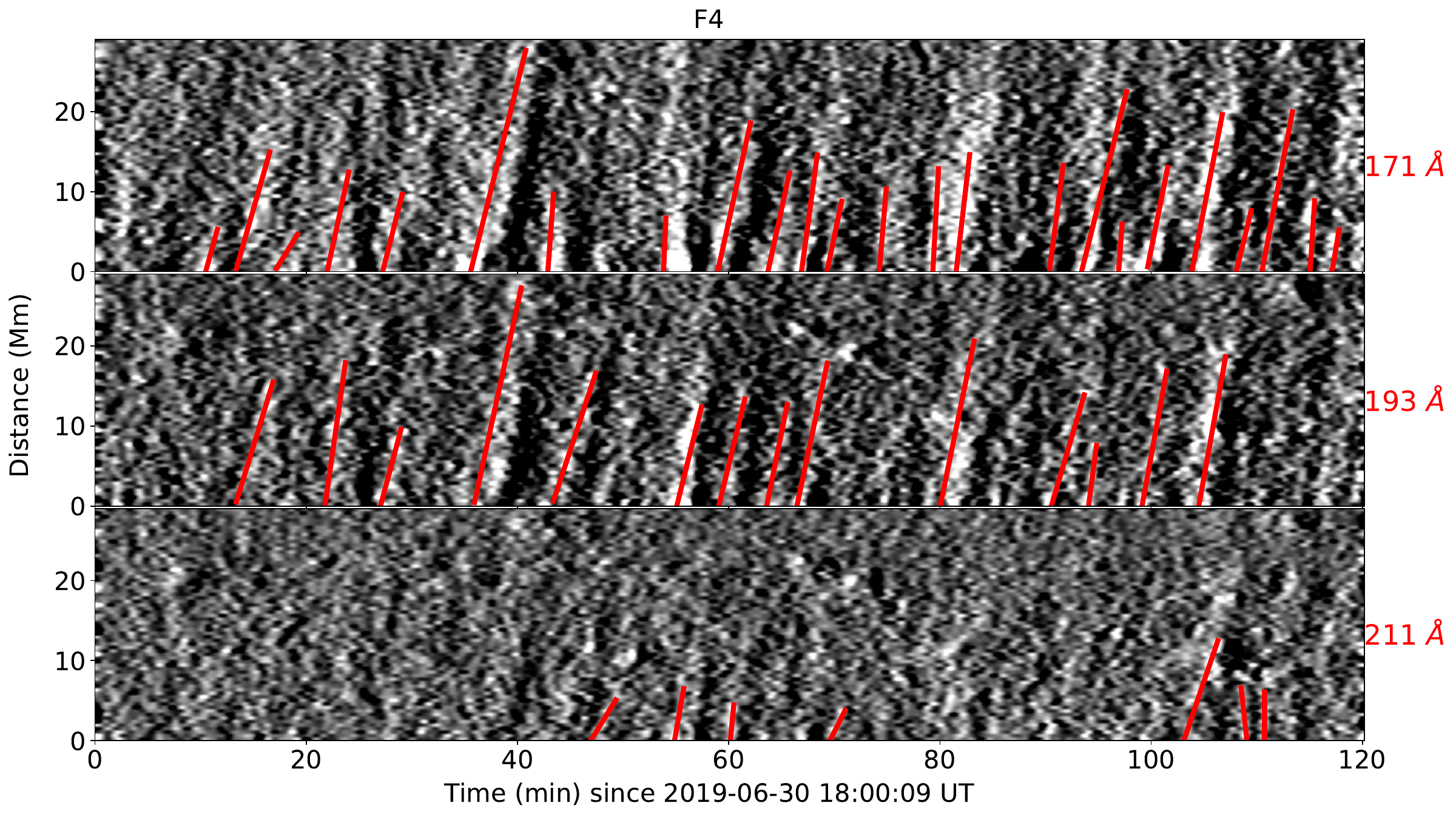}

    \caption{Panel \rm{I}: Space-time plot for the F3 time region. Panel \rm{II}: Space-time plot for the F4 time region of the plume. From top to bottom for each panel, we display the plot in 171~{\AA},193~{\AA}, and 211~{\AA} channels. Red lines are drawn to show outflow lines.}
    \label{fig:xt_plot56}
\end{figure}
\begin{figure}[htpb!]
    \centering
    \includegraphics[width=\linewidth]{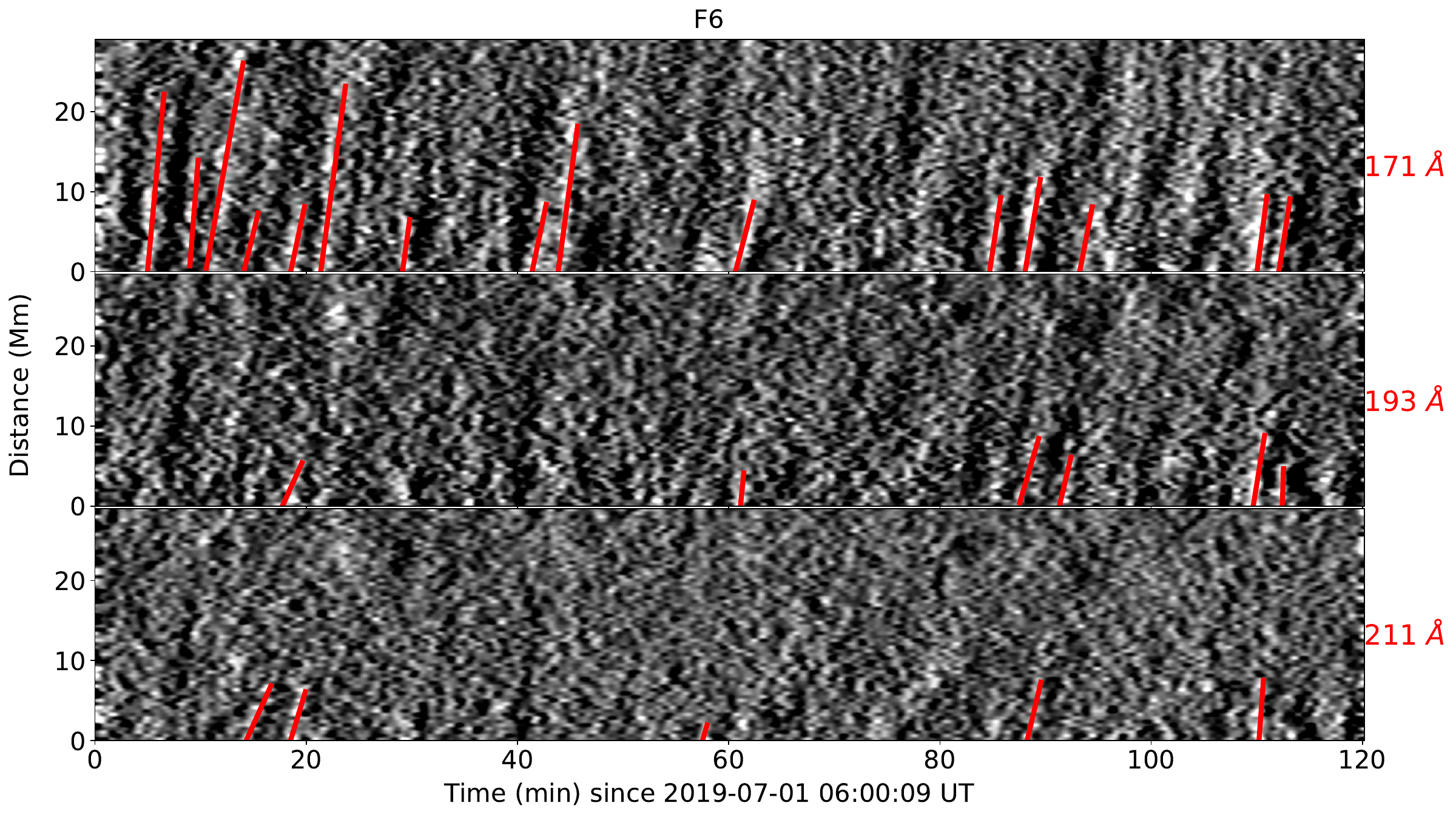}
    \includegraphics[width=\linewidth]{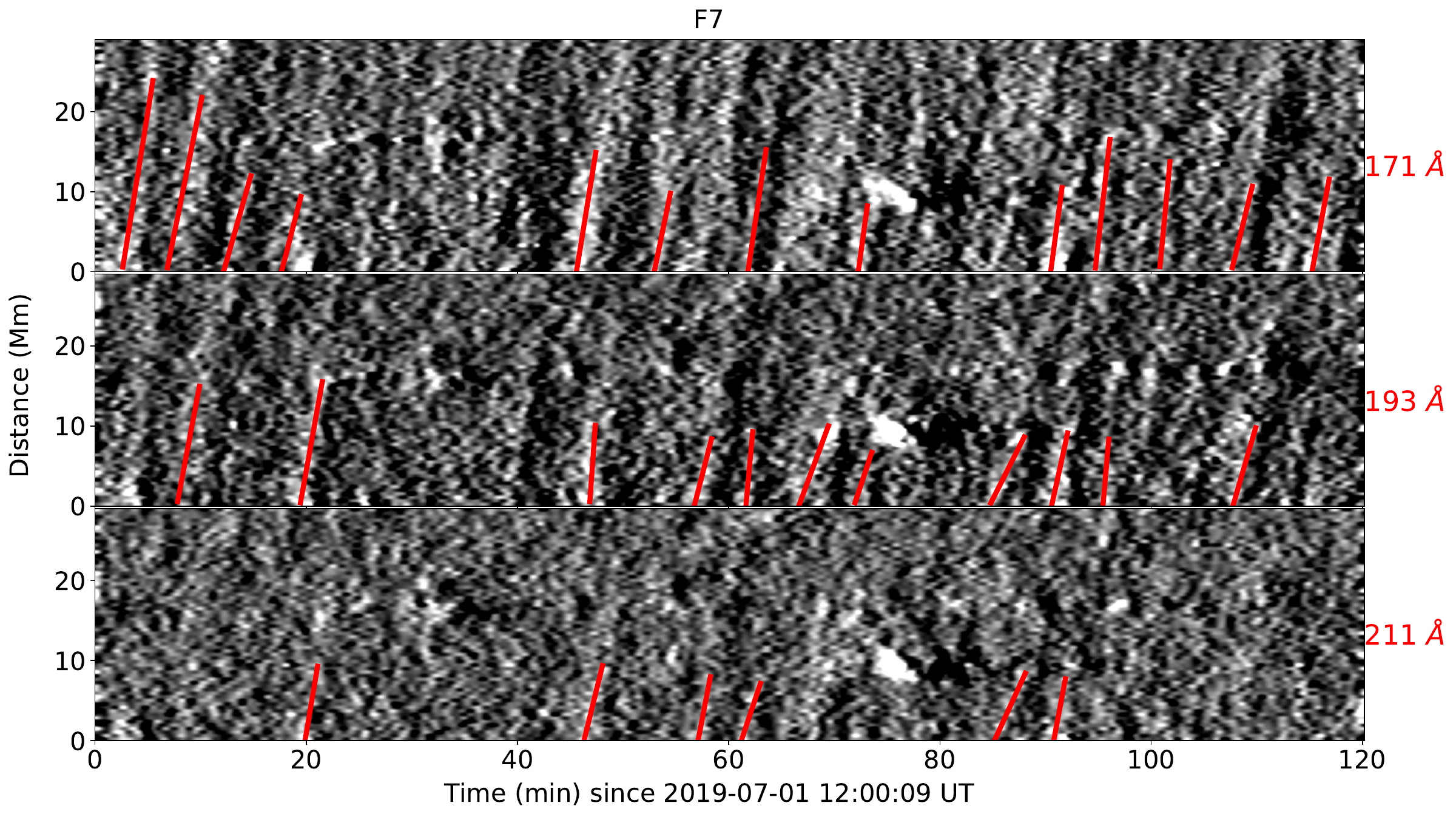}
    \caption{Panel \rm{I}: Space-time plot for the F6 time region. Panel \rm{II}: Space-time plot for the F7 time region of the plume. From top to bottom for each panel, we display the plot in 171~{\AA},193~{\AA}, and 211~{\AA} channels. Red lines are drawn to show outflow lines.}
    \label{fig:xt_plot78}
\end{figure}

\begin{deluxetable}{ccc|ccc}[H]
\tabletypesize{\footnotesize}
\tablewidth{0pt}
\tablecaption{Slits for temperature evolution. The position corresponds to the bottom left coordinate of the slit }
\label{tab:temp_evolution}
\tablehead{
\colhead{Date}   &   \colhead{Position}   &   \colhead{Angle}   &   \colhead{Date}   &   \colhead{Position}   &   \colhead{Angle}\\
\colhead{UT}   &   \colhead{(\arcsec,\arcsec)}   &   \colhead{Degree}   &   \colhead{UT}   &   \colhead{(\arcsec,\arcsec)}   &   \colhead{Degree}
}
\startdata
2019-06-30 00:00   &   [-195,870]   &   -8   &   2019-07-01 00:00   &   [-151,862]   &   -5\\
2019-06-30 01:00   &   [-195,870]   &   -8   &   2019-07-01 01:00   &   [-149,862]   &   -8\\
2019-06-30 02:00   &   [-193,870]   &   -8   &   2019-07-01 02:00   &   [-145,862]   &   -12\\
2019-06-30 03:00   &   [-189,870]   &   -10   &   2019-07-01 03:00   &   [-145,862]   &   -12\\
2019-06-30 04:00   &   [-187,870]   &   -8   &   2019-07-01 04:00   &   [-145,862]   &   -12\\
2019-06-30 05:00   &   [-186,870]   &   -10   &   2019-07-01 05:00   &   [-145,862]   &   -12\\
2019-06-30 06:00   &   [-183,868]   &   -10   &   2019-07-01 06:00   &   [-143,862]   &   -20\\
2019-06-30 07:00   &   [-180,869]   &   -12   &   2019-07-01 07:00   &   [-135,860]   &   -15\\
2019-06-30 08:00   &   [-178,868]   &   -7   &   2019-07-01 08:00   &   [-133,860]   &   -12\\
2019-06-30 09:00   &   [-176,868]   &   -14   &   2019-07-01 09:00   &   [-130,860]   &   -12\\
2019-06-30 10:00   &   [-176,868]   &   -14   &   2019-07-01 10:00   &   [-126,860]   &   -4\\
2019-06-30 11:00   &   [-173,868]   &   -14   &   2019-07-01 11:00   &   [-122,860]   &   -8\\
2019-06-30 12:00   &   [-170,868]   &   -8   &   2019-07-01 12:00   &   [-122,860]   &   -8\\
2019-06-30 13:00   &   [-168,868]   &   -7   &   2019-07-01 13:00   &   [-118,860]   &   -8\\
2019-06-30 14:00   &   [-168,865]   &   -15   &   2019-07-01 14:00   &   [-116,860]   &   -8\\
2019-06-30 15:00   &   [-168,865]   &   -18   &   2019-07-01 15:00   &   [-113,860]   &   -8\\
2019-06-30 16:00   &   [-165,865]   &   -15   &   2019-07-01 16:00   &   [-109,860]   &   -5\\
2019-06-30 17:00   &   [-165,865]   &   -15   &   2019-07-01 17:00   &   [-106,860]   &   -3\\
2019-06-30 18:00   &   [-165,865]   &   -15   &   2019-07-01 18:00   &   [-100,858]   &   -3\\
2019-06-30 19:00   &   [-165,865]   &   -15   &   2019-07-01 19:00   &   [-100,858]   &   -3\\
2019-06-30 20:00   &   [-158,862]   &   -16   &   2019-07-01 20:00   &   [-98,858]   &   -8\\
2019-06-30 21:00   &   [-158,862]   &   -2   &   2019-07-01 21:00   &   [-98,858]   &   -8\\
2019-06-30 22:00   &   [-155,862]   &   -2   &   2019-07-01 22:00   &   [-90,858]   &   -2\\
2019-06-30 23:00   &   [-151,862]   &   -5   &   2019-07-01 23:00   &   [-87,858]   &   -2\\
\enddata
\tablecomments{Angle is from north to east;
width=10\arcsec.
heght=45\arcsec.}
\end{deluxetable}



\end{document}